\documentclass[preprint2]{aastex}

\usepackage{natbib,amssymb}
\slugcomment{\apj\ accepted}

\shorttitle{MAMBO observations of SWIRE sources}
\shortauthors{Lonsdale et al.}

\def\rp{$r^\prime$}
\def\gp{$g^\prime$}
\def\rp{$r^\prime$}
\def\ip{$i^\prime$}
\def\msun{M$_{\odot}$}
\def\lsun{L$_{\odot}$}
\def\deg{$^{\circ}$}

\def\chandra {{\it Chandra}}
\def\xmm {XMM-{\it Newton}}
\def\spitzer {{\it Spitzer}}

\def\av {${\rm A_\mathrm{V}}$}

\def\kms{\ifmmode {\rm\,km\,s^{-1}}\else
    ${\rm\,km\,s^{-1}}$\fi}
\def\kmsMpc{\ifmmode {\rm\,km\,s^{-1}\,Mpc^{-1}}\else
    ${\rm\,km\,s^{-1}\,Mpc^{-1}}$\fi}
\def\ergAcm2{\ifmmode {\rm\,ergs\,cm^{-2}\,{\rm \AA}^{-1}}\else
    ${\rm\,ergs\,cm^{-2}\,\AA^{-1}}$\fi}
\def\cm2{\ifmmode {\rm\,cm^{-2}}\else
    ${\rm\,cm^{-2}}$\fi}
\def\ergcm2s{\ifmmode {\rm\,ergs\,cm^{-2}\,s^{-1}}\else
    ${\rm\,ergs\,cm^{-2}\,s^{-1}}$\fi}
\def\cgsdeg2{\ifmmode {\rm\,ergs\,cm^{-2}\,s^{-1}\,deg^{-2}}\else
    ${\rm\,ergs\,cm^{-2}\,s^{-1}\,deg^{-2}}$\fi}
\def\sqdeg{\ifmmode {\rm\,deg^{2}}\else
    ${\rm\,deg^{2}}$\fi}
\def\ergsHz{\ifmmode {\rm\,ergs\,s^{-1}\,Hz^{-1}}\else
    ${\rm\,ergs\,s^{-1}\,Hz^{-1}}$\fi}
\def\ergs{\ifmmode {\rm\,ergs\,s^{-1}}\else
    ${\rm\,ergs\,s^{-1}}$\fi}
\def\ergsA{\ifmmode {\rm\,ergs\,s^{-1}\,\AA^{-1}}\else
    ${\rm\,ergs\,s^{-1}\,\AA^{-1}}$\fi}
\def\WHz{\ifmmode {\rm\,W\,Hz^{-1}}\else
    ${\rm\,W\,Hz^{-1}}$\fi}
\def\WHzsr{\ifmmode {\rm\,W\,Hz^{-1}\,sr^{-1}}\else
    ${\rm\,W\,Hz^{-1}\,sr^{-1}}$\fi}
\def\ergscm2Hz{\ifmmode {\rm\,ergs\,cm^{-2}\,s^{-1}\,Hz^{-1}}\else
    ${\rm\,ergs\,cm^{-2}\,s^{-1}\,Hz^{-1}}$\fi}

\begin{document}


\title{MAMBO 1.2\,mm observations of luminous starbursts at $z\sim$2 in the SWIRE fields}


\author{Carol J. Lonsdale\altaffilmark{1,2}, Maria del Carmen Polletta\altaffilmark{2,3,4},
 Alain Omont\altaffilmark{4}, Dave Shupe\altaffilmark{5},
 Stefano Berta\altaffilmark{6,7}, Robert Zylka\altaffilmark{8}, 
 Brian Siana\altaffilmark{5}, 
 Dieter Lutz\altaffilmark{7}, 
 Duncan Farrah\altaffilmark{9}, Harding E. Smith\altaffilmark{2,16}, 
 Guilaine Lagache\altaffilmark{10},
 Carlos De Breuck\altaffilmark{11}, Frazer Owen\altaffilmark{12},
 Alexandre Beelen\altaffilmark{10}, Dan Weedman\altaffilmark{9}, Alberto Franceschini\altaffilmark{6}, 
 Dave Clements\altaffilmark{13}, Linda Tacconi\altaffilmark{7}, 
 Alejandro Afonso-Luis\altaffilmark{14},
 Ismael P\'erez-Fournon\altaffilmark{14}, Pierre Cox\altaffilmark{8},
 Frank Bertoldi\altaffilmark{15}}

\altaffiltext{1}{Infrared Processing \& Analysis Center, California Institute of Technology, 100-22, Pasadena, CA 91125, USA}
\altaffiltext{2}{Center for Astrophysics \& Space Sciences, University of California, San Diego, La Jolla, CA  92093--0424, USA}
\altaffiltext{3}{INAF-IASF Milano, via E. Bassini, 20133, Italy}
\email{polletta@lambrate.inaf.it}
\altaffiltext{4}{Institut d'Astrophysique de Paris, CNRS \& Universit\'e Pierre et Marie Curie, 98bis, bd. Arago, 75014 Paris, France}
\altaffiltext{5}{\spitzer\ Science Center, California Institute of Technology, 100-22, Pasadena, CA 91125, USA}
\altaffiltext{6}{Dipartimento di Astronomia, Universit\`a di Padova, Vicolo dell'Osservatorio 3, 35122 Padova, Italy}
\altaffiltext{7}{Max-Planck-Institut f\"ur extraterrestrische Physik, Postfach 1312, 85741 Garching, Germany}
\altaffiltext{8}{Institute de Radioastronomie Millimetrique, 300 rue de la Piscine, 38406 St. Martin d'Heres, France}
\altaffiltext{9}{Department of Astronomy, Cornell University, Ithaca, NY 14853, USA}
\altaffiltext{10}{Institut d'Astrophysique Spatiale, Universit\'e de Paris XI, 91405 Orsay Cedex, France }
\altaffiltext{11}{European Southern Observatory, Karl-Schwarzschild Strasse, 85748 Garching bei M\"unchen, Germany}
\altaffiltext{12}{National Radio Astronomy Observatory, P. O. Box O, Socorro, NM 87801, USA}
\altaffiltext{13}{Astrophysics Group, Imperial College London, Prince Consort Road, London SW7 2AZ, UK}
\altaffiltext{14}{Instituto de Astrofisica de Canarias, 38200 La Laguna, Tenerife, Spain}
\altaffiltext{15}{Max-Planck-Institut f\"ur Radioastronomie, Auf dem H\"ugel 69, 53121 Bonn, Germany}
\altaffiltext{16}{Deceased 2007 August 16.}

\begin{abstract} 

We report on--off pointed MAMBO observations at 1.2\,mm of 61
\spitzer-selected star-forming galaxies from the \spitzer\ Wide Area
Infrared Extragalactic Legacy survey (SWIRE). The sources are selected on
the basis of bright 24\,$\mu$m fluxes ($f_{24\,\mu m}$$>$0.4\,mJy) and of
stellar dominated near-infrared spectral energy distributions in order to
favor $z$$\sim$2 starburst galaxies. The average 1.2\,mm flux for the whole
sample is 1.5$\pm$0.2\,mJy. Our analysis focuses on 29 sources in the
Lockman Hole field where the average 1.2\,mm flux (1.9$\pm$0.3\,mJy) is
higher than in other fields (1.1$\pm$0.2\,mJy). The analysis of the
sources multi-wavelength spectral energy distributions indicates that they
are starburst galaxies with far-infrared luminosities from 10$^{12}$ to
10$^{13.3}$\,\lsun, and stellar masses of
$\sim$0.2--6$\times$10$^{11}$\msun.

Compared to sub-millimeter selected galaxies (SMGs), the SWIRE-MAMBO sources
are among those with the largest 24$\mu$m/millimeter flux ratios. The origin
of such large ratios is investigated by comparing the average mid-infrared
spectra and the stacked far-infrared spectral energy distributions of the
SWIRE-MAMBO sources and of SMGs. The mid-infrared spectra, available for a
handful of sources, exhibit strong PAH features, and a warm dust continuum.
The warm dust continuum contributes to $\sim$34\% of the mid-infrared
emission, and is likely associated with an AGN component. This constribution
is consistent with what is found in SMGs. The large 24$\mu$m/1.2\,mm flux
ratios are thus not due to AGN emission, but rather to enhanced PAH emission
compared to SMGs. The analysis of the stacked far-infrared fluxes yields
warmer dust temperatures than typically observed in SMGs. Our selection
favors warm ultra-luminous infrared sources at high-$z$, a class of objects
that is rarely found in SMG samples. Indeed SMGs are not common among bright
24\,$\mu$m sources (e.g. only about 20\% of SMGs have $f_{24\mu
m}>$0.4\,mJy). Our sample is the largest \spitzer-selected sample detected
at millimeter wavelengths currently available.
\end{abstract}


\section{Introduction}\label{intro}
 
After the discovery by IRAS of local Ultra-Luminous InfraRed Galaxies
(ULIRGs, with L$_{FIR}$ $\gtrsim$ 10$^{12}$ L$_{\odot}$) and follow-up
studies~\citep[see reviews by ][]{sanders96,lonsdale06}, the advent of
sensitive sub-mm bolometer arrays heralded a revolution in our understanding
of star formation at high-redshift. Extragalactic surveys with
SCUBA/JCMT~\citep{hughes98,scott02,mortier05,coppin06} and the Max Planck
Millimeter Bolometer (MAMBO) array~\citep{kreysa98} at the Institut de
Radioastronomie Millim\'etrique (IRAM) 30m telescope~\citep{greve04,voss06}
show that the number of dusty, infrared-luminous galaxies (high-$z$ ULIRGs,
commonly ``Sub-millimeter Galaxies'' or SMGs) at 2$\lesssim$z$\lesssim$3 is
several orders of magnitude higher than in the local universe and that these
distant SMGs contribute significantly to the universal star-formation
history~\citep[see e.g.][]{smail04}.  SMGs appear to be ``maximal
starbursts'' that form stars at the maximum global rate until a significant
fraction of their original large gas mass is converted into a massive
stellar system. Numerous follow-up studies indicate that SMGs are the most
likely objects to trace the major episodes of star formation leading to the
production of massive ($>2L^{*}$) ellipticals~\citep{greve05,tacconi06}.
This picture is consistent with the claim that evolved massive ellipticals
have formed the bulk of their stars in short ($<$\,2\,Gyr), intense
bursts~\citep{dunlop96,blakeslee99}. The evolution of SMGs into massive
elliptical galaxies is often attributed to a negative feedback mechanism
induced by Active Galactic Nuclei (AGN) activity~\citep{sanders88}. It is
speculated that as the super-massive black hole (SMBH) in these systems
reaches a certain mass and luminosity, the AGN will heat the surrounding
gas, and thus halt any further star-forming activity and SMBH growth. This
scenario is supported by deep X-ray observations of SMGs that show that some
SMGs contain fast growing SMBHs~\citep{alexander05,borys05}.

In this context, identifying a large sample of ULIRGs at the epoch where
these processes were taking place, i.e. $z\gtrsim$2, is particularly
important to understand the formation of the most massive galaxies,
investigate the contribution of SMGs in the star formation history of the
Universe, and the presence and role of AGN activity. These goals have up to
now been stymied by the limitations in mapping speed and sensitivity of the
current sub-mm bolometers. The total number of known SMGs from all
blank-field surveys with any instrument is less than five hundred.  Their
number and luminosity function are known from MAMBO/SCUBA counts, but remain
uncertain at both low and high luminosities. Moreover, in order to quantify
the relative importance of BH growth, stellar mass assembly, and the
negative feedback induced into the host galaxies by these processes, it is
necessary to separate any starburst and AGN power in these systems. Such an
analysis requires a multi-wavelength coverage and a good characterization of
their spectral energy distributions (SEDs) and spectra. For these reasons,
numerous efforts are being carried out to collect large samples of
ULIRGs at $z\simeq$2 with well sampled SEDs and spectra.

\section{\spitzer\ observations of sub-millimeter galaxies}\label{lit_smg}

With the advent of wide area \spitzer\ surveys, new breakthroughs in the
study of SMGs are now possible as the statistics of the high redshift ULIRG
population increases and the power sources in SMGs are quantified. In
particular, the \spitzer\ Wide Area Infrared Extragalactic
Survey~\citep[SWIRE~\footnote{http://swire.ipac.caltech.edu/swire/swire.html};
][]{lonsdale03}, the largest ($\sim$50\,deg$^2$) of the \spitzer\ suite of
extragalactic surveys, with its sensitivity is able to detect many sources
found in blank-field sub-mm/mm surveys. The SWIRE 5$\sigma$ limits are 4.2,
7.5, 46, 47\,$\mu$Jy in the 4 IRAC~\citep{fazio04} bands at 3.6, 4.5, 5.8,
and 8.0\,$\mu$m, respectively, and 0.21, 18, and 108 mJy in the 3
MIPS~\citep{rieke04} bands at 24, 70, and
160\,$\mu$m~\citep{surace05}\footnote{The given SWIRE 5$\sigma$ limits are
valid for the LH field. They can differ by up to a factor of 1.2 in the
other fields.}. Many previously known SMGs lie in the SWIRE fields and a
large fraction of them are detected in 2 or more IR bands~\citep[see
statistics below and][]{clements08}. 

In order to estimate which fraction of SMGs would be detected by \spitzer\
to the SWIRE sensitivities, we gathered from the literature a sample of 90
SMGs with published \spitzer\ data, or that lie in the SWIRE
fields~\citep{greve04,ivison04,egami04,frayer04a,frayer04b,chapman04b,borys05,pope06}.
This sample is not meant to be complete or exhaustive, but it includes a
large number of objects that are representative of sub-mm selected SMGs
(`classical SMGs' hereinafter). 

From an analysis of the SMGs with $f_{1.2mm} > 2.5$$\,$mJy or S$_{850\mu m}$
$>$7$\,$mJy (22 `bright SMGs'), we expect the large majority to be detected
in at least 2 SWIRE bands.  Ten (45\%) of these 22 `bright SMGs' are
detected above the SWIRE 24$\mu$m $\sim$5--6$\sigma$ flux density limit
($\sim$250$\mu$Jy), 7 of them (32\%) are brighter than 400$\mu$Jy, 20 (91\%)
are detected at both 3.6 and 4.5$\mu$m above the SWIRE 5$\sigma$ limits,
18\% are detected at 5.8$\mu$m and 27\% at 8.0$\mu$m\footnote{The
\spitzer\ 70 and 160$\mu$m imaging sensitivity is such that extremely few
high redshift ULIRGs are detectable in large area surveys such as SWIRE.}.

The density of bright ($f_{1.2mm} > 2.5$$\,$mJy or S$_{850\mu m}$
$>$7$\,$mJy) SMGs is $\sim$230 per sq.\ deg.~\citep{voss06,coppin06}, which
means that in $\sim$50 sq.\ deg.\ SWIRE would detect with IRAC about
$\sim$10,000 sources that are typical of those found in sub-mm surveys, and
$\sim$5,000 at 24$\mu$m ($>$3,000 brighter than 400$\mu$Jy), which is 10
times larger than all existing MAMBO-SCUBA surveys put together. Appreciable
SWIRE detectability rates are also found for the sub-mm fainter SMGs
(3.2$<$S$_{850\mu m}$$<$7$\,$mJy): 36\% detectable at 24$\mu$m (14\%
brighter than 400$\mu$Jy); 86\% at 3.6 and 4.5$\mu$m, 5\% at 5.8$\mu$m and
9\% at 8$\mu$m, so many thousand more sub-mm fainter SMGs are also to be
found among SWIRE sources, including many 24$\mu$m-detected sources. 
Moreover, \spitzer\ offers the opportunity to investigate possible selection
biases inherent to sub-mm surveys in favor of objects dominated by
cool dust.

The identification of high-$z$ ULIRGs among the large SWIRE population of
galaxies is not trivial, however, because they are faint in the optical and
their optical-MIR SEDs can be similar to those of more quiescent star
forming galaxies. Moreover, only long wavelength data (far-IR and mm/sub-mm)
and reliable redshifts can confirm the extreme luminosities of the ULIRG
candidates, but only 0.4\% of the SWIRE population is detected at $\lambda
\geq$\,70$\mu$m, and most of these ULIRG candidates are at $z$$<$1 or too
faint in the optical to be followed-up spectroscopically.

X-ray, radio, IR and CO
studies~\citep{alexander05,chapman05,greve05,valiante07,menendez07} indicate
that the far-IR (FIR) emission of the bulk of SMGs is dominated by
starbursts, but a large fraction of them also host weak AGNs. 
\spitzer\ holds a key to this question because the diagnostic provided by
the MIR broadband SEDs and IRS~\citep{houck04} spectra can discriminate warm
AGN-dominated emission. Indeed, using IRS spectra, \citet{pope08} constrain
the AGN contribution to the MIR emission of SMGs to $<$30\%. However,
\spitzer\ alone cannot determine the FIR luminosity or dust temperature of
these systems because most of them are not detected at 70 and 160$\mu$m.

Therefore, millimeter or sub-millimeter observations are needed to identify
these objects and directly measure the long wavelength emission and estimate
their FIR luminosities. The combination of SWIRE and MAMBO offers the
possibility to explore the role of ULIRGs at $z\sim$2 and determine their
power sources.

We have thus undertaken a systematic observing program with MAMBO at 1.2\,mm
to search for luminous ULIRGs among SWIRE sources. The results from the
first set of observations from this program are presented here.

This work is organized as it follows. \S~\ref{samp_sel} describes the
selection of the best candidates from the four northern SWIRE fields. Their
observations with MAMBO are presented in \S~\ref{obs}. The spectral energy
distributions (SEDs) of all sources and photometric redshift estimates are
presented in \S~\ref{zphot_seds}. Millimeter and 24\,$\mu$m fluxes,
broad-band SEDs, and colors of the selected sources are compared with those
of sub-mm selected galaxies in \S~\ref{comp}. FIR luminosities and star
formation rates (SFRs) are estimated by fitting greybody models with dust
temperatures typically measured in SMGs, and presented in \S~\ref{fir_seds}.
Since the majority of the sources are not detected with MIPS at 70 and
160\,$\mu$m, we measure average values from stacked MIPS images. The stacked
FIR fluxes at 70 and 160$\mu$m, combined with the average 24\,$\mu$m and
1.2\,mm fluxes, are used to characterize the FIR emission of these sources,
and to compare it with the FIR emission of SMGs in \S~\ref{fir_seds}.
Stellar masses, near-IR (NIR) luminosities and star-formation timescales are
discussed in \S~\ref{masses}. The presence of AGN activity is investigated
using different tracers, MIR spectra, X-ray, and radio observations in
\S~\ref{agn_contr}. The possible nature of the selected sources and their
relation with classical SMGs are discussed in \S~\ref{discussion}. Finally,
our results are summarized in \S~\ref{summary}. Most of the analysis
described here is applied to the sub-sample of sources from the Lockman Hole
field. We only report the observations and resulting data on the sub-samples
from the other fields. This choice is due to the fact that the Lockman Hole
sub-sample is larger and better observed than those in the other fields. The
sub-samples from the other fields do not allow us to carry out an accurate
statistical analysis. We adopt a flat cosmology with H$_0$\,=\,71\kmsMpc,
$\Omega_{M}$\,=\,0.27 and $\Omega_{\Lambda}$\,=\,0.73~\citep{spergel03}.

\section{Sample selection}\label{samp_sel}

The most sensitive of the \spitzer\ MIPS bands for identification of distant
ULIRGs is the 24$\mu$m band. Therefore, we have searched for extreme ULIRGs
among SWIRE sources with $f_{24}>250{\mu}$Jy, corresponding to
$\gtrsim$5$\sigma$. Most systems above this $f_{24}$ limit and z$>$1 are
expected to be ULIRGs. In addition distant ULIRGs will be faint at optical
wavelengths, therefore we limited our overall z$>$1 ULIRG samples to sources
with \rp$>$23. This is a brighter cut than adopted in previous works
targeting extreme 24\,$\mu$m sources~\citep[e.g.][]{houck05,yan05}. For our
purpose, we prefer to adopt a brighter optical limit because a fainter cut
can reject ULIRGs with colors like some local well known examples, e.g.; an
Arp\,220-like system ($z$=0.018 and L(IR) = 1.45$\times$10$^{12}$\,\lsun) at
$z$=1 with an observed $f_{24}$=250$\mu$Jy would have a luminosity of
L$_{IR}=10^{12.6}$\lsun\ and an observed magnitude $r^{\prime}$=23.4, while
an Arp\,220-like system with L$_{IR}=10^{13}$\lsun\ at z=1 would have an
observed $r^{\prime}$=22.4. 

The selection based on bright 24\,$\mu$m and faint \rp-band fluxes yields
consequently sources with large $f_{24\,\mu m}/f_r$ ratios, typically
$\gtrsim$100. Bright 24$\mu$m \spitzer\ samples with large IR/optical flux
ratios (i.e., $f_{24\,\mu m}/f_R>$500) tend to be dominated by AGN light in
the MIR~\citep{houck05,yan05,sajina07,brand06,dey08}. Moreover MAMBO observations
of these have shown that they are weak at 1.2\,mm compared to sub-mm
selected SMGs~\citep{lutz05}. An additional selection criterion was thus
applied to disfavor AGN-dominated and favor star-forming-dominated systems.
The additional selection criterion is based on the evidence in the IRAC
bands for the $\sim$ 1.6$\mu$m rest-frame Planck spectrum peak from evolved
stars. Note that IRAC was in part designed for photometric selection of
galaxies at $z$=1.5--3 displaying this feature~\citep{simpson99,sawicki02}.
Such a peak is not observed in templates of local AGN-dominated sources such
as Mrk\,231, IRAS\,19254$-$7245~\citep{berta03}, or in type 1
QSOs~\citep{elvis94}. On the other hand, the SED of sources displaying such
a peak can be well fitted by local star formation-dominated templates, such
as M\,82, and Arp\,220~\citep{silva98}.

Furthermore, we have selected sources where this peak lies in the 5.8$\mu$m
band in order to favor sources at $z$=1.5--3. We refer to such sources as
`5.8\,$\mu$m-peaker' (or `b3' for bump in the third IRAC channel). Formally
a source is defined `5.8\,$\mu$m-peaker' if satisfies the following
condition: $f_{3.6}<f_{4.5}<f_{5.8}>f_{8.0}$ where $f$ is the flux density
in $\mu$Jy and $f_{8.0}$ can also be an upper limit (typically at
5$\sigma$).

We have previously demonstrated that this selection method is effective in
finding starburst-dominated objects at
$z$=1.5--3~\citep{weedman06a,berta07,farrah08}. IRS MIR spectroscopic
observations of a similar sample of 32 SWIRE sources displaying a rest-frame
NIR bump peaking at 4.5\,$\mu$m demonstrate that they are star-forming
dominated~\citep{farrah08}. Most of their spectra exhibit prominent PAH
emission features consistent with a starburst origin of their MIR emission.
Similarly, \citet{weedman06a} find that all SWIRE sources with a NIR bump at
5.8\,$\mu$m, display PAH emission features in their IRS spectra.  Moreover,
8 of the sources selected here were observed in the MIR by the IRS
instrument on \spitzer~\citep[6 in the LH; ][ and 2 in the EN2 field,
Lonsdale, in prep.]{weedman06a}, and their spectra display strong PAH
features as expected from starburst-dominated emission. The IRS observations
yield spectroscopic redshifts that, in all but one case, confirm the
`5.8\,$\mu$m-peakers' photometric redshift selection method.  The agreement
between photometric and IRS redshifts is discussed further in
\S~\ref{zphot_seds}~\citep[see also][]{farrah08}. Based on these
considerations and studies, we infer that the combination of a strong
24\,$\mu$m flux and a peak in the IRAC bands is a good indication of the
presence of PAH emission at $z\sim$2, and thus of a significant, and likely
dominant, starburst contribution to the NIR and MIR emission. Although in
some cases an AGN component in the MIR can be present, its contribution is
never dominant compared to the starburst component~\citep{berta07b}.

In order to further increase the chance of selecting $z\sim$2 star-forming
galaxies, we derived photometric redshifts for all source candidates (for a
detailed description of the procedure see \S~\ref{zphot_seds}).

We have selected in this manner 1023 `5.8\,$\mu$m-peaker' ULIRG candidates
with $f_{24}>$ 250$\mu$Jy, \rp$>$23, and photometric redshift in the range
1.5--3, from the SWIRE's largest field (11\,deg$^2$), the Lockman Hole (LH).
We restricted the search to the region in the LH with available optical
imaging ($\sim$9.5\,deg$^2$). At similar redshifts and for fixed SED shape
the most luminous objects will be the brightest in flux. Therefore, to cull
from this large sample the most luminous candidates we selected the
brightest 24\,$\mu$m sources, i.e. $f_{24}>$400$\mu$Jy, among all
`5.8\,$\mu$m-peakers', and with $\geq$5$\sigma$ detections in the
first 3 IRAC bands. There are 520 sources with $f_{24}>$400$\mu$Jy,
corresponding to 55 sources per square degree. We have also required
unambiguous (unique or not detected counterpart) source identification at
all optical-MIPS wavelengths in the images, so that source confusion would
not affect the SED interpretation.  Our selection is likely to strongly
favor sources with strong 7.7$\mu$m PAH features in 24$\mu$m band in the
$z$=2 range.

\begin{figure*}
\epsscale{2.0}
\plotone{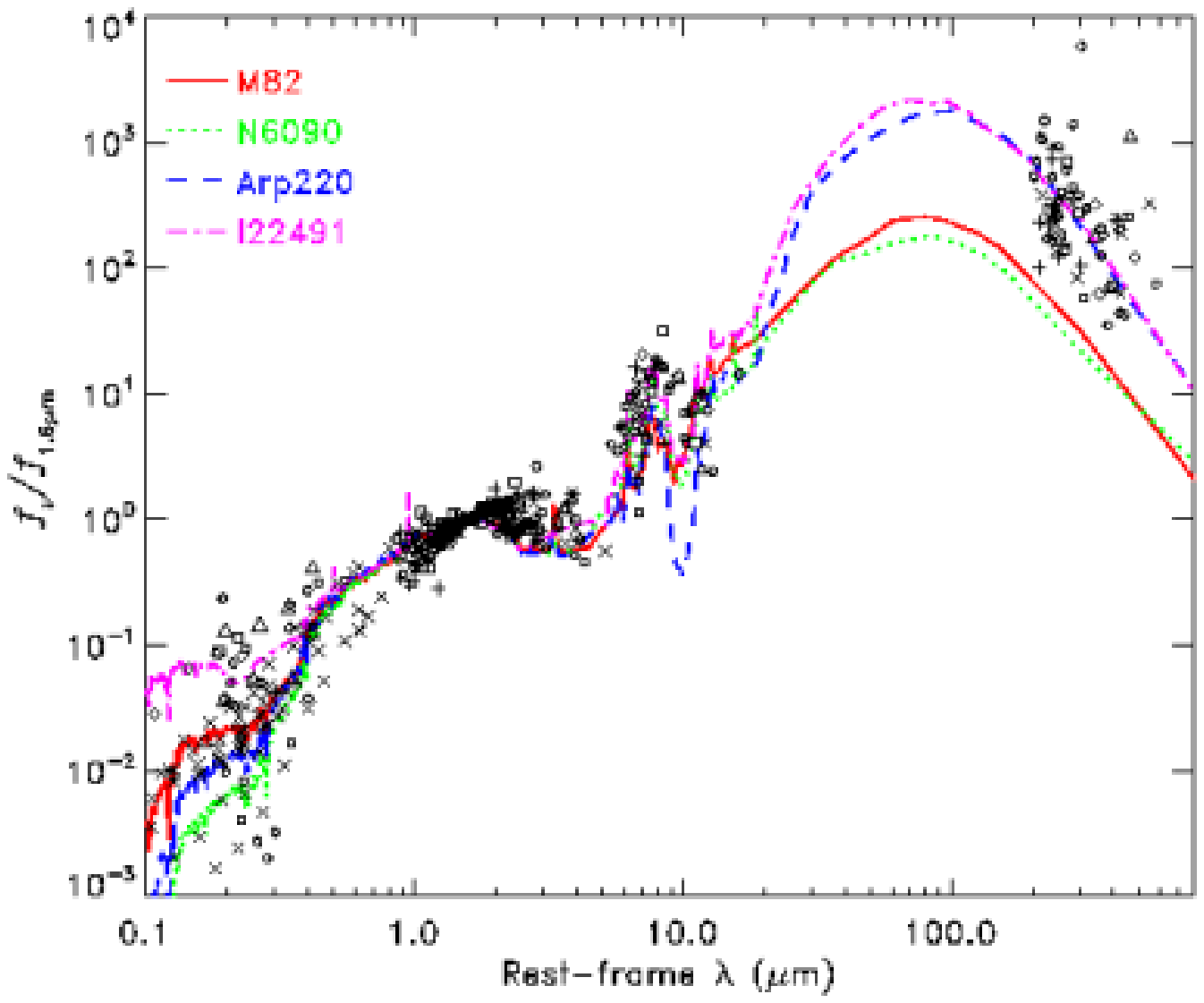}
\caption{Rest-frame SEDs normalized at 1.6\,$\mu$m of sub-mm selected galaxies from the
literature (open circles: HDFN sources from~\citet{pope06}, crosses: GOODS
sources from~\citet{borys05}, squares: Lockman Hole sources, and triangles:
ELAIS-N2 sources from~\citet{chapman04b}, diamonds: Lockman Hole sources
from~\citet{greve04}, stars: FLS sources
from~\citet{frayer04a}, and plus signs: GOODS sources from~\citet{egami04}.
The curves represent various starburst galaxy templates: M\,82 (solid red), NGC\,6090 (dotted
green), Arp\,220 (dashed blue), and I22491 (dot-dashed magenta).}
\label{smg_seds_lit}
\end{figure*}

\begin{figure*}
\epsscale{2.0}
\plotone{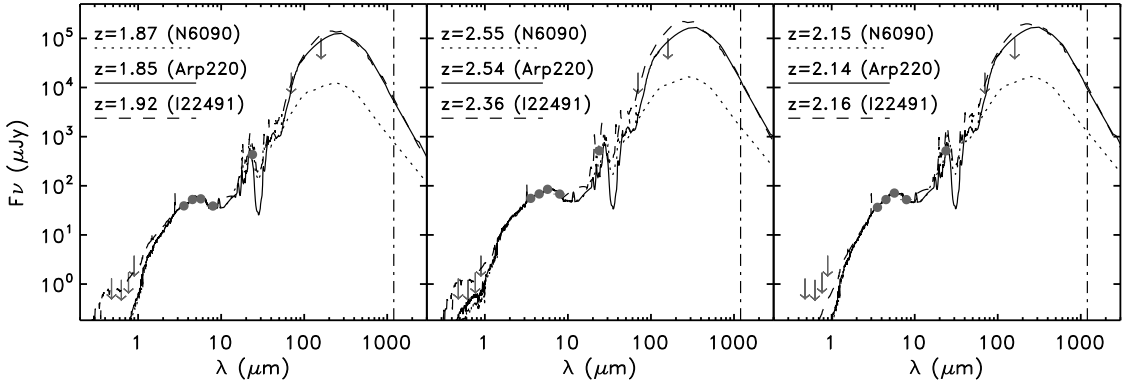}
\caption{Fits of 3 SWIRE/MAMBO sources with the 3 starburst templates used to predict
1.2mm flux densities, Arp 220 (solid curve), NGC 6090 (dotted curve), and
I22491 (dashed curve). Downward arrows represent 5$\sigma$ upper limits. The
vertical dot-dashed line represents the 1.2\,mm wavelength.}
\label{two_seds_fits}
\end{figure*}

Finally we made estimates of the expected 1.2\,mm flux densities by fitting
templates of various local starbursts and ULIRGs to the optical ($Ugriz$)
and infrared (3.6--24$\mu$m) bands. Since most of the candidates are not
detected at 70 or 160$\mu$m, their FIR through mm spectral shape is
unconstrained. The MAMBO flux predictions therefore rest on the adoption of
a long wavelength SED template. To select appropriate templates we build the
rest-frame SEDs of $\sim$30 SMGs with known redshifts and available IRAC and
24$\mu$m fluxes, and normalized them at rest-frame 1.6$\mu$m.  We then
selected templates of local starburst-dominated ULIRGs which pass
approximately through the midst of the distribution of 850$\mu$m and 1.2mm
points for these objects, as shown in Figure~\ref{smg_seds_lit}. Most of the
SMGs have FIR SEDs similar to Arp\,220 or IRAS 22491$-$1808 (I22491
hereinafter). The main difference between the Arp\,220 and I22491 templates,
a starburst dominated and a composite starburst+AGN system, respectively,
resides at optical and MIR wavelengths. However, since with some extinction
the optical emission of I22491 resembles that of Arp\,220, and in the MIR we
do not have enough spectral coverage to favor one template with respect to
the other, we decided to fit each candidate source with an Arp\,220
template. In addition we decided to also fit a template with a much lower
FIR/NIR flux ratio than Arp\,220, to obtain minimum estimates of the
predicted 1.2mm flux for each source.  For this we chose NGC\,6090, a local
starburst with a similar SED shape to the well known starburst galaxy M\,82.
Examples of these two fits are shown for 3 sources in
Figure~\ref{two_seds_fits}, where it can be seen that the two templates
produce similar photometric redshifts from the IRAC data, but the
predictions for the bolometric luminosity, and for the 1.2\,mm flux density
differ by a factor of about 10.

To select the best candidates for observations with MAMBO, we averaged the
predicted 1.2mm flux density for each source from both templates and ordered
the sources according to decreasing predicted averaged 1.2\,mm flux.  While
this procedure may not predict the 1.2mm flux of an individual source
accurately, because the actual long wavelength SED is unconstrained, it does
provide a robust way to prioritize them by decreasing the expected 1.2\,mm
flux under the assumption that they have similar SED shapes, and taking into
account the sources redshift.

The same selection procedure was applied to other three SWIRE fields,
XMM-LSS (XMM), ELAIS-N1 (EN1) \& ELAIS-N2 (EN2). A description of the fields
sizes and multi-wavelength coverage can be found in~\citet{surace05}. The
top 61 sources from the 4 SWIRE fields, (LH, XMM, EN1 \& EN2), were observed
with MAMBO at 1.2\,mm. The largest number of sources was selected in the LH
field, 29 sources, compared to 17 in EN1, 9 in EN2, and 6 in XMM. The
predicted 1.2\,mm fluxes (averaged on all templates) range from 1.4 to 25
mJy. The list of selected sources, coordinates and optical-infrared-mm
fluxes are reported in Tables~\ref{tab_mm} and~\ref{tab_optir}.  

Note that because of the uncertainties on the IRAC fluxes, especially at 5.8
and 8.0\,$\mu$m, which are less sensitive bands than those at shorter
wavelengths, a `5.8\,$\mu$m-peaker' ($z\sim$2.2--3.2) can become
`4.5\,$\mu$m' or `8.0\,$\mu$m-peakers' (`b2', or `b4') in different versions
of the data processing. Indeed, some of the sources selected here, that were
previously classified `5.8\,$\mu$m-peakers' based on a previous version of
the SWIRE catalog used during the selection process, show a peak at shorter
(`b2'; $z\sim$1.5--2.2) or longer wavelengths (`b4'; $z\geq$3.2) or a flat
IRAC SED (`fl') in the latter SWIRE catalogs that is used in this work. This
change is observed in 14 out of 61 selected objects (see last column of
Table~\ref{tab_mm}).

\section{MAMBO observations and results}\label{obs}

Observations were spread over the pool observing sessions at the IRAM 30m
telescope in Fall/Winter 2005/2006, using the 117 element version of the
MAMBO array~\citep{kreysa98} operating at a wavelength of 1.2\,mm (250 GHz). 
On the first run in late October - early November 2005 we observed most of
the Lockman sample (29 sources), and the 6 XMM sources. During the Winter
run, mainly early March 2006, the observations of the Lockman sample were
completed and exploratory observations of the 17 EN1 and 9 EN2 sources were
carried out.

We used the standard on-off photometry observing mode, chopping between the
target and sky at 2 Hz, and nodding the telescope every 10 or 20\,s. On-off
observations were typically obtained in blocks of 6 scans of 16 or 20
10s-subscans each, and repeated in later observing nights. The atmospheric
transmission was intermediate with
$\tau$(1.2\,mm) at zenith between 0.1 and 0.4. The absolute flux calibration
was established by observations of Mars and Uranus, resulting in a flux
calibration uncertainty of about 20\%.

On average, the noise of the channel used for point-source observations was
about 35--40 mJy/$\sqrt{t}$/beam, where $t$ is the exposure time in seconds,
consistent with the MAMBO time estimator for winter conditions. In order to
identify a maximum number of strong sources with flux density
$\sim$3--5\,mJy within the allocated time (57h, i.e. $\sim$38h of
integration time), we adopted a two-step procedure: a first exploratory
phase generally aiming at an rms $\sim$1.0--1.2\,mJy; then longer
integrations on sources showing hints of detection in order to reach
rms\,$\approx$\,0.6--0.7\,mJy and achieve 3--5$\sigma$ detections of sources
stronger than $\sim$2--3\,mJy. However, we were able to apply such a
complete procedure only to the LH sample (29 sources), which had the best
selection and was the only one observable in good conditions (during night
or early morning) both in Fall and Winter runs. The observations of the
sources of the three other fields (EN1 and EN2, XMM; 32 sources) were more
or less limited to the exploratory phase.

A major change occurred in the 30m-telescope control system in 2005.  MAMBO
pool observations in our October-early November session were carried out
with a well known system which had been in use for years. A new
system, just installed before the Winter pool session, brought some problems
such as stronger accelerations of the telescope during nodding with possible
effects on tracking accuracy and temperature stability of the helium
cryostat in some circumstances; anomalous noise in some MAMBO channels for a
few cases; etc. These problems induced some delays and loss of time, but
they were practically all well under control for most of our observations in
February-March 2006. The data were reduced with standard procedures in the
MOPSIC package developed by R.~Zylka. Great care was taken to check the
quality of the data obtained with the new control system, and we discarded
the data from about ten scans displaying obvious problems such as anomalous
noise in some channels, as well as $\sim$3\% of the data taken in periods
when we are not sure that the problems with the new system were well under
control.

Table~\ref{tab_mm} lists the resulting 1.2\,mm flux densities and their
statistical uncertainties.  In the more completely studied Lockman sample
(29 sources, Table~\ref{tab_mm}), more than 40\% of the sources (12) show a
signal at $>$2.5\,$\sigma$, and fluxes
$\gtrsim$\,2\,mJy~\citep[corresponding to $\sim$5\,mJy at 850\,$\mu$m for
$z\sim$2, see ][]{greve04}. Eight of them have 1.2\,mm fluxes in the
3--5\,mJy range, with S/N\,$\gtrsim$\,4--5, which guarantees the robustness
of the detections. However, it is possible that the flux densities of the
detected sources are slightly overestimated by the Malmquist bias, by an
amount that probably does not exceed 10\% on average. The average flux
density of the 29 LH sources is 1.88$\pm$0.28\,mJy (see
Table~\ref{tab_stat}), corresponding to $\sim$5\,mJy at 850\,$\mu$m at
$z\sim$2).

As quoted, the remaining 32 sources, from EN1 \& EN2 and XMM, were not
observed as completely. However, the average strength of the sources in the
other fields is significantly weaker (1.14$\pm$0.20\,mJy) than in the
Lockman sample. The fraction of sources showing at least a 2.5$\sigma$
signal is only 25\% (8 sources), there is practically no source stronger
than 3\,mJy, and the average of the 23 other undetected sources is only
0.60$\pm$0.17 mJy. Possible reasons for such a behavior are either a
pointing problem caused by the system changes implemented in 2005, or a bias
introduced by the slight different selection criteria (see
\S~\ref{mm_fluxes}).

Table~\ref{tab_stat} summarizes the information about the number of sources
showing a signal at various levels and the average values for two
sub-samples and the totality of the 61 sources.

\section{Photometric redshifts and SED fits}\label{zphot_seds}

In order to classify the spectral energy distributions (SEDs) and estimate
photometric redshifts of the sources in the sample, optical and IR data
($\leq$24\micron) are combined and fitted with various galaxy templates. The
SEDs are fitted using the {\sf Hyper-z} code~\citep{bolzonella00}. {\sf Hyper-z}
finds the best-fit by minimizing the $\chi^2$ derived from the comparison of
the observed SED and expected SEDs at various redshifts, and using a
template library and the same photometric system. The effects of dust extinction
are taken into account by reddening the reference templates according to a
selected reddening law. We use the prescription for extinction measured in
high-redshift starbursts by~\citet{calzetti00}.  We limit the additional
extinction \av\ to be less than 2.0 mag and include templates of highly
extinguished objects. We used two template libraries, one with 105 templates
of starburst galaxies~\citep{chary01}, and one with 8 templates including 2
late-spirals, 5 starbursts, and 1 composite (starburst+AGN) template
covering the wavelength range between 1000\AA\ and 1000\micron. The
starburst templates correspond to the SEDs of NGC\,6090, NGC\,6240,
Arp\,220, IRAS\,20551$-$4250, and I22491~\citep{silva98,berta05}.
The composite (AGN+SB) template corresponds to the SED of the Seyfert 2
galaxy IRAS 19254$-$7245 South~\citep[I19254;][]{berta03}. A description of
the library and of the method can be found in~\citet{polletta07}.

We fit the SEDs using the data up to 24$\mu$m, including upper limits.
Although the optical upper limits place some constraints on the acceptable
fits, for these optically faint objects the photometric redshifts are based
primarily on the IRAC bands. The photometric redshifts are therefore limited
in accuracy by (1) the small number of available photometric bands (3--4
IRAC bands, MIPS[24], and up to a maximum of 5 optical bands among $Ugriz$),
and (2) the fact that the main spectral feature available for fitting (the
peak in the stellar spectrum) is broad and detected through broad filters.
Therefore our photometric redshifts are expected to have uncertainties of
$\pm$0.5 in redshift or higher, even for the highest SNR sources. In order
to illustrate the range of acceptable solutions and photometric redshifts,
we report multiple solutions if they give similar $\chi^2$. The fits were
also done without including the 24$\mu$m measurements. Since the comparison
with spectroscopic redshifts was slightly better when those data were
included, we decided to adopt the results obtained using also the 24$\mu$m
data.  The inclusion of 24\,$\mu$m data in the fitting procedure can yield
better results, if templates including dust emission, like the libraries
used here, are employed~\citep[see e.g.][]{polletta07}.

\begin{figure*}
\epsscale{2.0}
\plotone{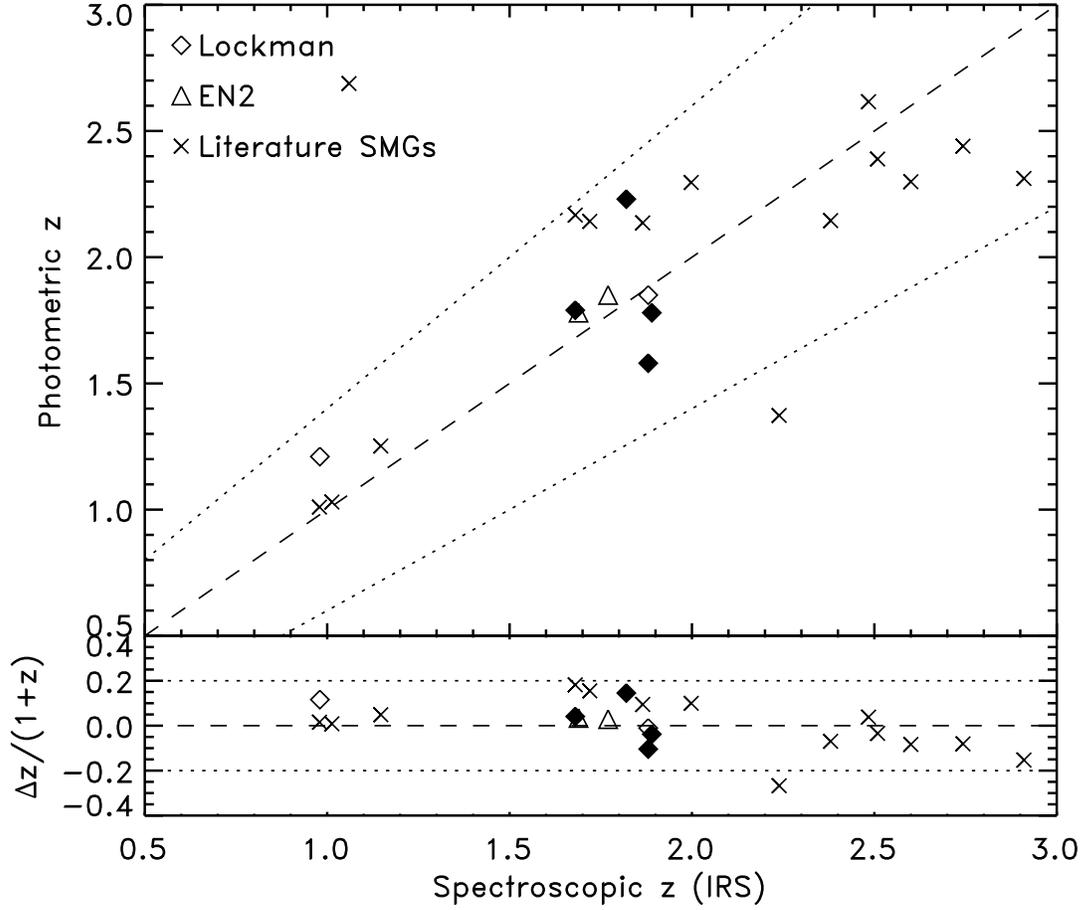}
\caption{Comparison between photometric and spectroscopic redshifts from
 IRS~\citep[Lonsdale et al., in prep.][]{weedman06a} of the 8 SWIRE/MAMBO
 sources with available spectroscopic redshifts (diamond and triangles). 
 Dotted lines represent 20\% agreement in (1+$z$). Full symbols represent
 sources with $f_{1.2\,mm}$$>$2$\sigma$, and empty symbols sources with
 $f_{1.2\,mm}$$<$2$\sigma$.  Diamonds represent LH sources, and triangles
 sources in the other three SWIRE fields. Crosses represent 15 sources from the
 literature sample of SMGs with available spectroscopic redshifts, a
 24\,$\mu$m detection, and exhibiting a peak in the IRAC bands.}
  \label{zphot_zspe}
\end{figure*}

The reliability and accuracy of the photometric redshifts can be estimated
using the available spectroscopic data.  Spectroscopic redshifts from the
IRS are available for 8 sources and are listed in
Table~\ref{tab_fits}~\citep[][ Lonsdale et al., in prep]{weedman06a}. 
We define the fractional error $\Delta z$, the systematic mean error
$\overline{\Delta z}$, the 1$\sigma$ dispersion
$\sigma_z$, and the rate of catastrophic outliers, defined as the fraction
of sources with $\left| \Delta z\right| >$ 0.2. $\Delta z$ is defined as:
\begin{equation}
\Delta z = \left(\frac{z_{phot}-z_{spec}}{1+z_{spec}}\right) 
\end{equation} 
and
\begin{equation} 
\sigma_z^2 = \sum \left(\frac{z_{phot}-z_{spec}}{1+z_{spec}}\right)^2/N 
\end{equation} 
with N being the number of sources with spectroscopic redshifts. The
systematic mean error, $\overline{\Delta z}$, is 0.03. The $rms$ of
$\overline{\Delta z}$, $\sigma_z$ is 0.08, and there are no outliers with
$\left| \Delta z\right| >$ 0.2. The comparison between the spectroscopic and
final photometric redshifts is shown in Figure~\ref{zphot_zspe} for the 8
sources. There is a good agreement for 5 sources, and for 3 sources
there is a difference $\gtrsim$0.05 in Log(1+$z$). In order to better asses
the reliability of our estimates, we applied the same technique to a sample
of 24\,$\mu$m detected SMGs from the literature with known spectroscopic
redshift and with at least 3 detections in the IRAC bands consistent with
the presence of a peak at either 4.5\,$\mu$m or 5.8\,$\mu$m. We found 15
sources that satisfy these criteria out of the 90 SMGs for which we
collected data from the literature (see \S~\ref{intro}). Note that most of
the sources from the literature do not benefit from observations from as
many bands as the SWIRE-MAMBO sample. After combining the SWIRE-MAMBO and
the literature sample, $\overline{\Delta z}$=0.03, $\sigma_z$=0.19, and we
find 2 outliers ($\left| \Delta z\right| >$ 0.2), corresponding to $\sim$9\%
of the sample. The results obtained for the literature sample confirm the
accuracy of our estimates on the SWIRE-MAMBO sample.

SED fits for all the SWIRE sources in the sample are provided in the
Appendix and the best-fit templates and photometric redshifts are listed in
Table~\ref{tab_fits}.

\begin{figure*}
\epsscale{2.0}
\plotone{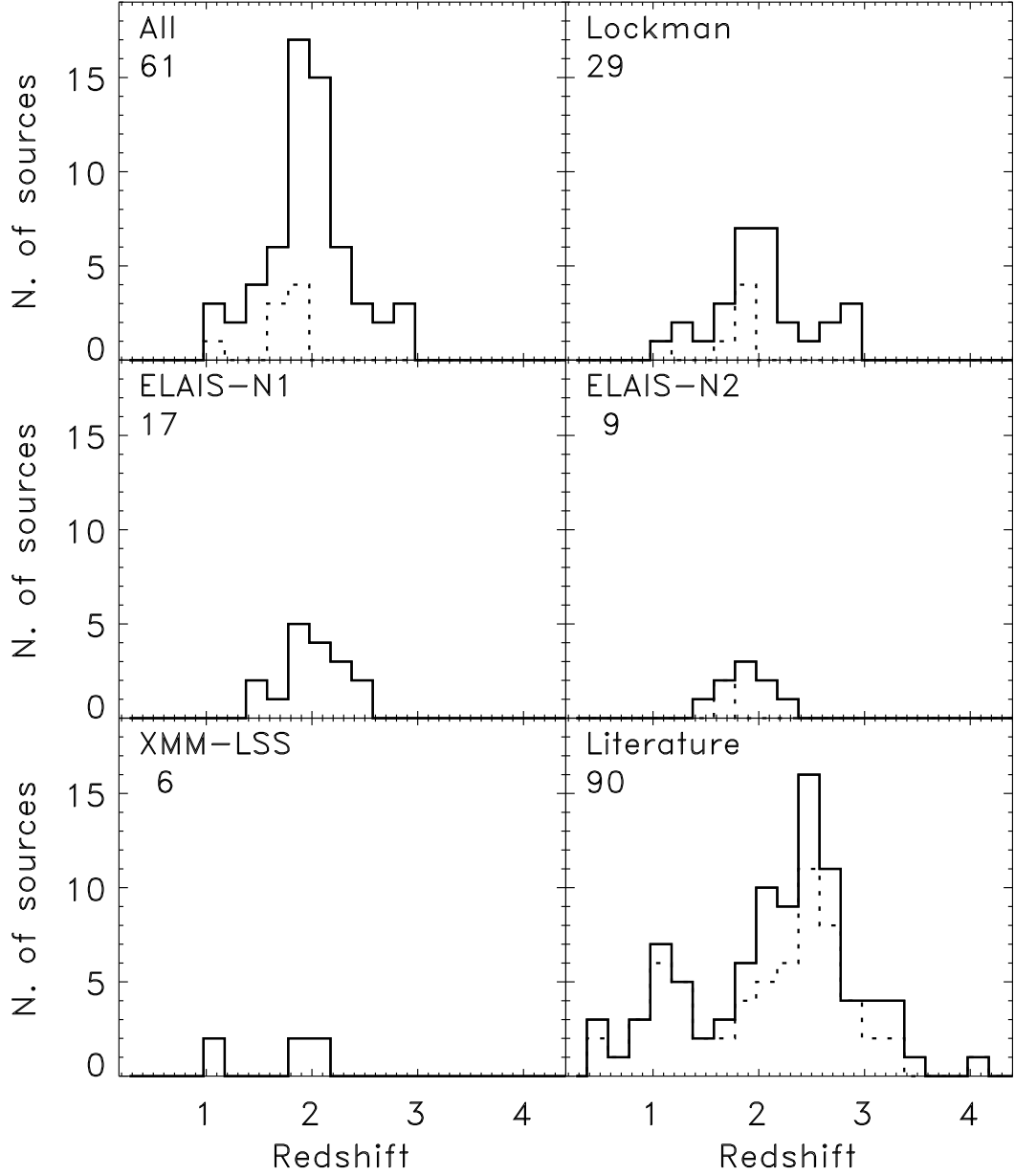}
\caption{Redshift distribution for all observed sources. The distribution of
spectroscopic redshifts is shown with a dotted line. From top to bottom and
from left to right, all MAMBO/SWIRE sources, those in the Lockman Hole,
ELAIS-N1, ELAIS-N2, and XMM-LSS, and from the
literature~\citep{egami04,borys05,chapman04b,greve04,pope06}. The number of
sources is annotated on the upper left corner of each panel.}
  \label{histo_z}
\end{figure*}

The redshift distribution of the sample, based on the photometric redshifts
(solid line) and on the spectroscopic redshifts (dotted line) when available, is
shown in Figure~\ref{histo_z}. The distribution peaks at
$z\simeq$2.0, which corresponds to about the minimum redshift for a 5.8\,$\mu$m
peaker. The $z$ peak of our sources is slightly lower than the redshift
distribution of sub-mm-selected
sources~\citep{frayer04b,egami04,borys05,chapman05,greve04}. Note that the
difference could be even larger because of the possible overestimate of
z$_{phot}$ in case, for example, of a shift of the peak of the stellar bump
due to an AGN contribution~\citep[e.g. a `b2' can appear as a `b3' and its
redshift be overestimated by $\Delta z$=1, see e.g. ][]{berta07,daddi07}.

\section{General properties of the SWIRE-MAMBO sources}\label{comp}

\subsection{Millimeter and 24\,$\mu$m fluxes}\label{mm_fluxes}

\begin{figure}
\epsscale{1.0}
 \plotone{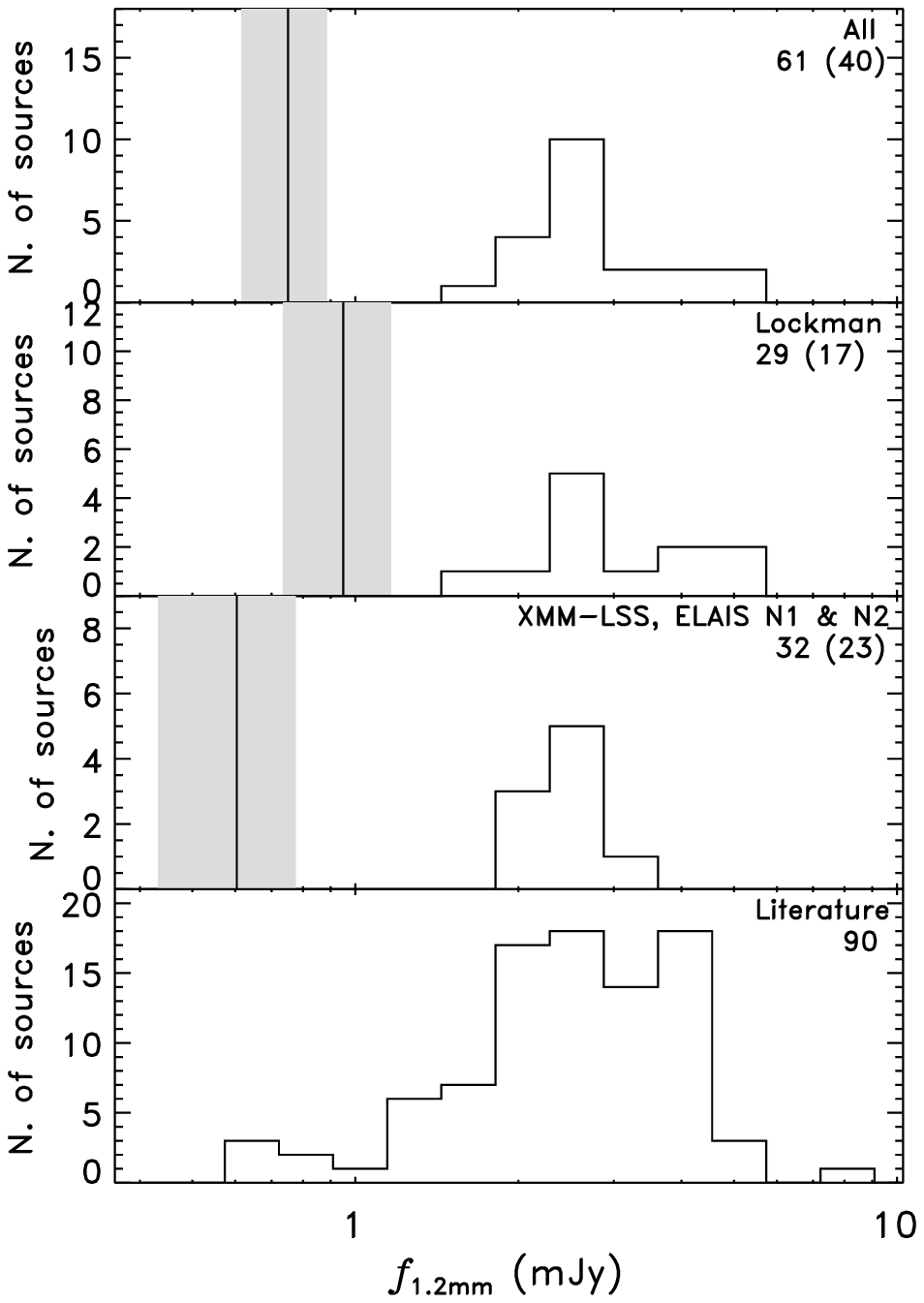}
 \caption{1.2mm flux distribution of sources with
 $f_{1.2\,mm}$$>$2$\sigma$ (histogram). The mean and mean standard deviation of the
 sources with $f_{1.2\,mm}$$<$2$\sigma$ are shown as a vertical solid line and a
 gray shaded area. From top to bottom, all MAMBO/SWIRE sources, those in the
 Lockman Hole, those in the the other fields, and those in the
 literature~\citep{frayer04b,egami04,borys05,chapman04b,greve04,pope06}. The
 total number of plotted sources is annotated and in parenthesis the number
 of sources with $f_{1.2\,mm}$$<$2$\sigma$ is given.}
\label{histo_mm}
\end{figure}

The distribution of observed 1.2\,mm flux densities (separated in
sources with $f_{1.2mm}$$\geq$2$\sigma$ and $<$2$\sigma$) is illustrated in
Figure~\ref{histo_mm}. The fraction of $>$2$\sigma$ sources, 41\% (24\% at
4$\sigma$), and the large average 1.2\,mm flux observed in our samples
(1.5$\pm$0.2\,mJy), and especially in the more thoroughly observed LH sample
(1.9$\pm$0.3\,mJy), show that our criteria are quite successful in selecting
sub-millimeter-bright galaxies from MIR samples.

The XMM, EN1 \& EN2 sample show systematically lower fluxes compared to the
LH sample. Although the LH sample was more thoroughly observed than the
other samples, with typical exposure times $>$1.5 hours, compared to
$\sim$40 min for the other sources, the origin of the different detection
rate and average fluxes is more likely due to intrinsic properties of the
sources or of the fields, e.g. cosmic variance, than to the observational
parameters.  The most plausible explanation is that our selection introduced
different biases in the selected samples possibly due to the different
optical coverage or to the analysis carried out on the SEDs and on the
images during the source selection process. The latter was indeed more
extensive in the LH field.

In Figure~\ref{histo_mm}, we also show the distribution of 1.2\,mm fluxes of
the 90 SMGs selected from the literature (see \S~\ref{lit_smg}).  In case of
lack of flux measurement at 1.2\,mm for the SMGs from the literature, we
estimate it from the flux at 850\,$\mu$m by applying the following equation:
\begin{equation}
Log(f_{1.2\,mm}) = -0.36+0.98\times Log(f_{850\,\mu m})
\label{smm_mm}
\end{equation}
where $f_{1.2\,mm}$ and $f_{850\,\mu m}$ are the flux densities at 1200 and
850\,$\mu$m, respectively. This relationship is derived from a sample of 21
SMGs for which measurements at both 850 and 1200\,$\mu$m are
available~\citep{greve04}. 

Based on Figure~7 in~\citet{voss06} and the 850\,$\mu$m source counts from
SHADES~\citep{coppin06}, the expected number of SMGs with
$f_{1.2mm}>$5\,mJy, 4\,mJy, and 3\,mJy in the 9\,deg$^2$ LH field is
$\sim$250, 800, and 2,000, respectively. On the other hand, we have
identified only 2, 3, and 4 sources brighter than 5\,mJy, 4\,mJy, and
3\,mJy, respectively - out of 29 24\,$\mu$m bright sources in the 9\,deg$^2$
LH field. The small number of mm-bright sources in our sample and in the
sample in~\citet{lutz05} suggests that very luminous millimeter sources are
not common among sources with bright 24\,$\mu$m fluxes ($f_{24\mu
m}>$400\,$\mu$Jy).

\begin{figure}
\epsscale{1.0}
 \plotone{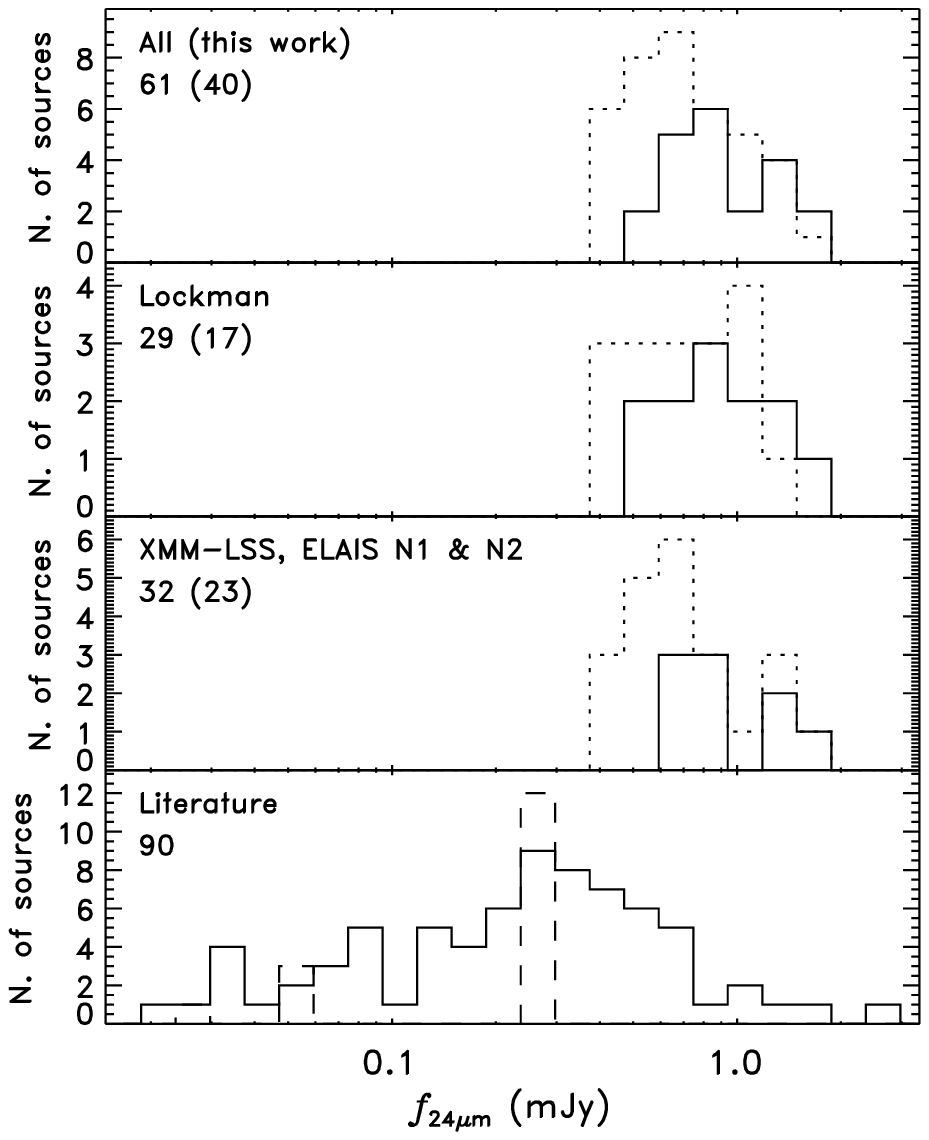}
 \caption{24\,$\mu$m flux distribution of sources with
 $f_{1.2\,mm}$$>$2$\sigma$ (solid line) and with $f_{1.2\,mm}$$<$2$\sigma$
 (dotted line). From top to
 bottom, all MAMBO/SWIRE sources, those in the Lockman Hole, those in the
 the other fields, and those in the
 literature~\citep{frayer04b,egami04,borys05,chapman04b,greve04,pope06}. 
 The dashed histogram represents the literature sources that are not
 detected at 24\,$\mu$m for which we assumed a 5$\sigma$ upper limit
 (250\,$\mu$Jy for 12 sources in the SWIRE fields, and 26\,$\mu$Jy for 1
 GOODS SMG, and 506\,$\mu$Jy for 3 more GOODS SMGs). The total number of
 sources is annotated on the upper left corner in each panel. The value 
 in parenthesis represents the number of sources with
 $f_{1.2\,mm}$$<$2$\sigma$.}
\label{histo_24um}
\end{figure}

In Figure~\ref{histo_24um}, we show the distribution of 24\,$\mu$m fluxes of
the SWIRE-MAMBO sample and of the SMGs from the literature. All SWIRE-MAMBO
sources have 24\,$\mu$m greater than 400\,$\mu$Jy as imposed by the
selection criteria. The 24$\mu$m fluxes of our sample range from
0.4 to 1.5 mJy, with a median value of 0.76\,mJy and a mean value of
1.09$\pm$1.06\,mJy. The SWIRE-MAMBO sources with $f_{1.2\,mm}$$<$2$\sigma$
show a similar range of 24\,$\mu$m fluxes, but fewer of the faintest
24\,$\mu$m sources are detected by MAMBO.  Figure~\ref{histo_24um} clearly
illustrates that the majority of SMGs are characterized by 24\,$\mu$m fluxes
smaller than 400$\mu$Jy. More specifically, out of the 90 SMGS from the
literature, 20 (22\%) have 24\,$\mu$m fluxes greater than 400\,$\mu$Jy.
Therefore, about 78\% of classical SMGs would be missed by our selection
criteria. Indeed, more than half of SMGs at $z\sim$2 are not detected at the
SWIRE sensitivity at 24\,$\mu$m, $\sim$250\,mJy (and even at 5.8 or
8.0\,$\mu$m, see~\S~\label{intro}). The lowest 24\,$\mu$m flux in the SMGs reaches
21\,$\mu$Jy, but there are 16 sources that are not detected, 4 in GOODS
($f_{24\mu m}<$26--50\,$\mu$Jy) and the others in the SWIRE fields
($f_{24\mu m}<$250\,$\mu$Jy). Since our sources are at similar redshifts and
show similar mm fluxes than SMGs, they either represent a subset of 24$\mu$m
bright SMGs ($\sim$20\%) or a different class of SMGs with enhanced MIR
emission. This question will be further investigated in the following
sections.

\subsection{Comparison with other \spitzer\ samples}

A number of works have recently been published about the MIR properties of
SMGs. They are based on hundreds of sources detected both with \spitzer\
from 3.6 to 24\,$\mu$m and with SCUBA at 850\,$\mu$m or with MAMBO at
1.2\,mm (see \S~\ref{lit_smg}). Most of these sources have been identified
in deep fields mapped both with SCUBA or MAMBO and
\spitzer~\citep{egami04,ivison04,greve04,chapman05,borys03,pope05,pope06,ivison07,yun08},
or in pointed \spitzer\ observations at the position of known SCUBA or MAMBO
sources~\citep{charmandaris04,frayer04a,frayer04b,ashby06,valiante07,menendez07,pope08}.
All these studies focus on sub-mm or mm selected galaxies, while our
\spitzer\ selection aims at sources bright at 24\,$\mu$m.  However, it is
interesting to note that our sources satisfy the selection criteria based on
IRAC colors developed to select SMG candidates among the IRAC
population~\citep[see eq. 1 in][]{yun08}.

Only a few other studies have a similar approach to ours in characterizing
the millimeter properties of \spitzer\ mid-IR selected sources.
\citet{lutz05} study a sample of bright ($f_{24\mu m}\gtrsim$1\,mJy) sources
with high 24$\mu$m/optical flux ratios, and $f_{24\mu m}$/$f_{8\mu m}$ flux
ratios to favor star-forming galaxies, but which ended up to be largely
contaminated by AGNs~\citep{yan05,yan07,sajina07}. Others study
AGN-dominated samples in the SWIRE survey~\citep[][Perez-Fournon et al., in
prep.]{polletta08c}. The SWIRE-MAMBO sample differs from those studies
because special care was applied to favor star forming galaxies and remove
AGNs that usually dominate the population of 24\,$\mu$m bright and optically
faint sources~\citep{brand06}. Since the presence of a bump in the
rest-frame NIR is due to stellar emission, our selection disfavors sources
with dominant AGN emission in the MIR. Our selection thus makes this sample
unique with respect to other \spitzer-selected samples that have been
followed-up at sub-mm and mm wavelengths

\subsection{Spectral energy distributions}

\begin{figure*}
\epsscale{2.4}
\plottwo{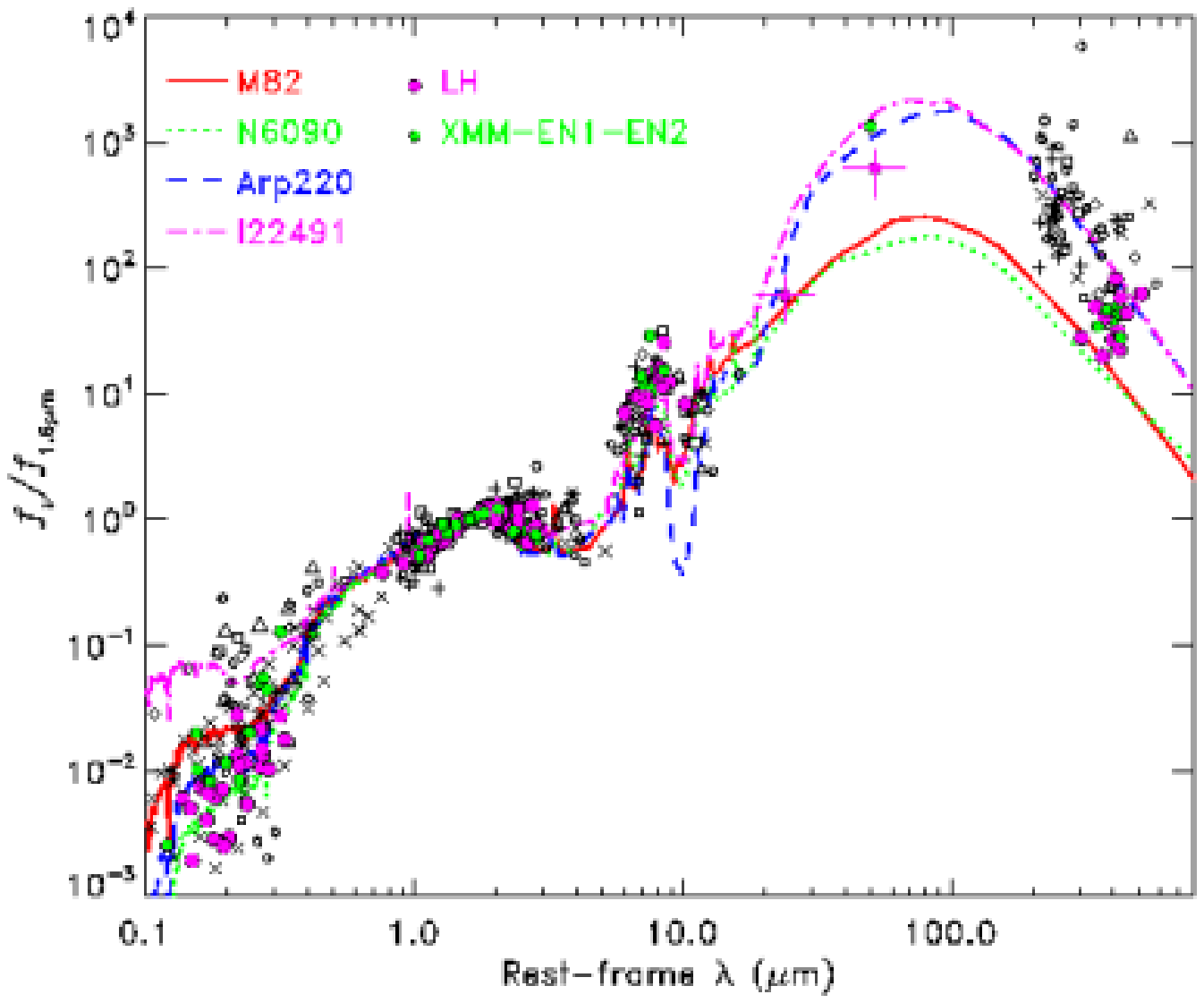}{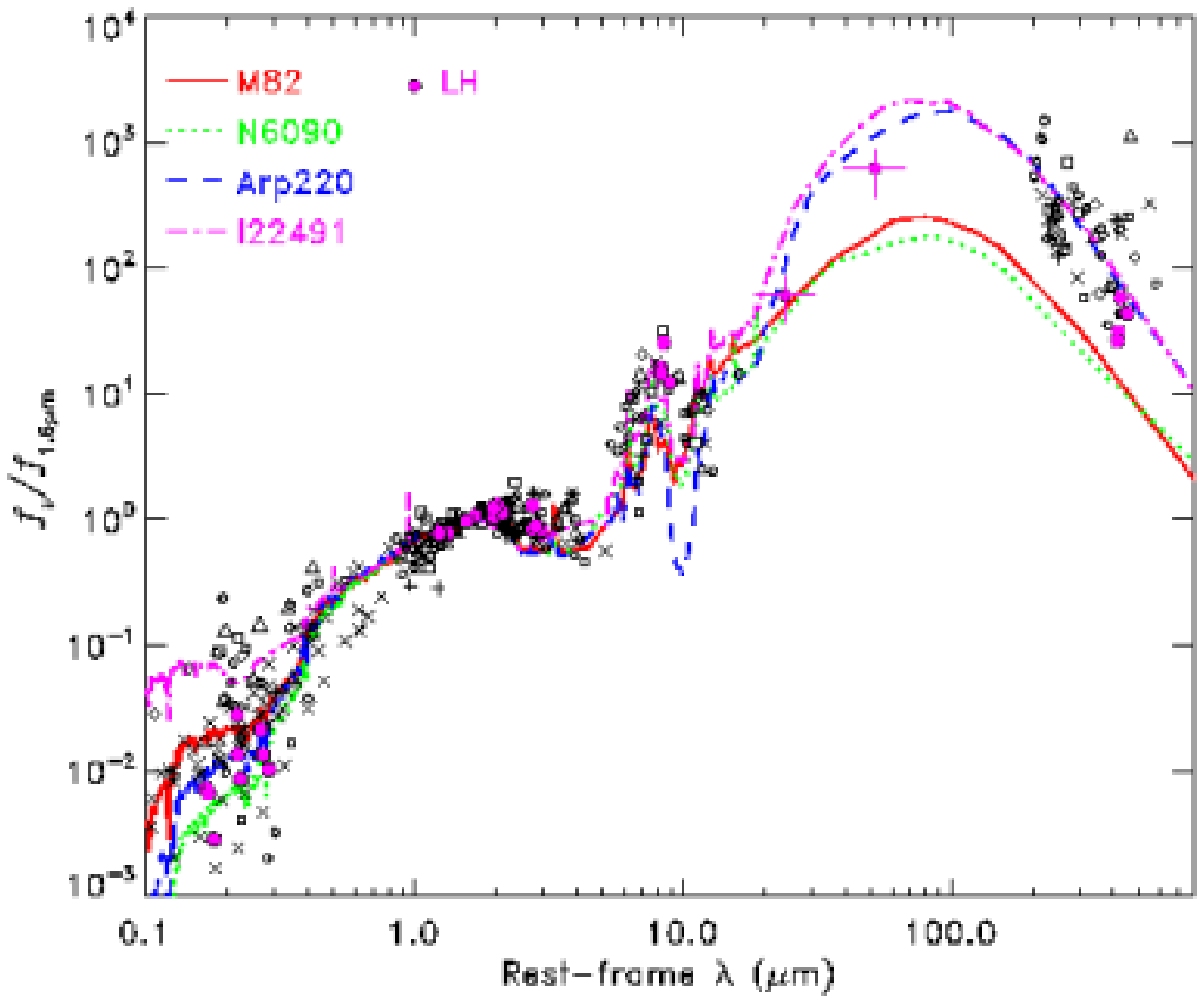}
\caption{Rest-frame SEDs of SMGs from the literature (black symbols as in
figure~\ref{smg_seds_lit}) and of the SWIRE/MAMBO sources with 1.2\,mm
fluxes $>$2$\sigma$ (full circles, red: Lockman Hole, green: XMM-LSS,
ELAIS-N1, or ELAIS-N2) normalized at 1.6\,$\mu$m. The 70 and 160\,$\mu$m
fluxes correspond to the normalized stacked fluxes of all SWIRE/MAMBO
sources with 1.2\,mm fluxes $>$2$\sigma$. The curves represent starburst
templates: M\,82 (solid red), NGC\,6090 (dotted green), Arp\,220 (dashed
blue), and I22491 (dot-dashed magenta). All sources are shown on the left
panel and only sources with a spectroscopic redshifts are shown on the right
panel.}
\label{smg_seds_lit_swire}
\end{figure*}

The spectral energy distributions (SEDs) of our sources are compared with
those of sub-mm selected SMGs in Figure~\ref{smg_seds_lit_swire}. Only
sources with spectroscopic redshifts are shown in the right panel.  All SEDs
are shown in the rest-frame and normalized at 1.6$\mu$m in the rest frame. 
The SWIRE sources are predominantly high in the 24$\mu$m band at the
location of the rest-frame 7.7$\mu$m PAH emission feature, and low in the
rest-frame FIR compared to the SMGs from the literature. The FIR emission of
the SWIRE sources is in between that expected for the typical starburst
galaxies as M\,82, and NGC\,6090, and a ULIRG like Arp\,220,
after normalizing them at the 1.6\,$\mu$m in the rest-frame. 

Figure~\ref{smg_seds_lit_swire} shows clearly the selection bias that is
present in our sample: by selecting the 24$\mu$m-brightest sources with a
peak at 5.8\,$\mu$m, we have strongly favored sources at $z\sim$2 with large
NIR and MIR luminosities. The classical SMG sample is naturally more biased
towards the most mm-bright systems. The location of our sources in
Figure~\ref{smg_seds_lit_swire} with respect to various template tracks
confirms that their SEDs are intermediate between M\,82 and Arp\,220.

\subsection{Colors and nature of the sources}\label{colors}

\begin{figure*}
\epsscale{2.0}
\plotone{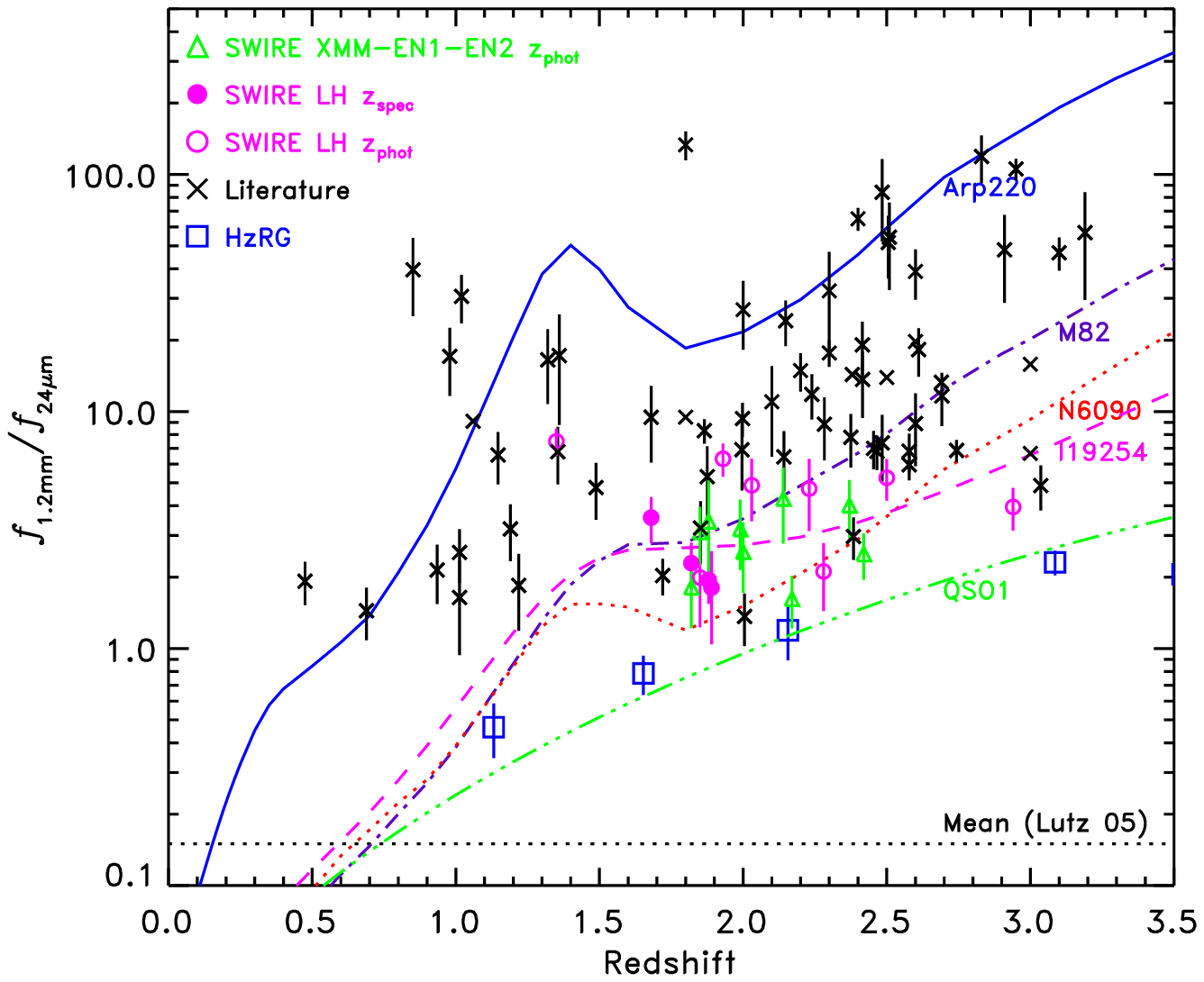}
\caption{Observed flux ratios, $f_{1.2\,mm}/f_{24\,\mu m}$, as a function of redshift
for SMGs from the literature (black crosses), for SWIRE/MAMBO
sources with $f_{1.2\,mm}$$>$2$\sigma$ (circles: LH, triangles: other fields), and for high-$z$ radio
galaxies~\citep[blue squares ][]{seymour07}. Expected values for various
starburst and AGN templates are also shown: Arp\,220 (solid blue), M\,82
(dot-dashed purple), NGC\,6090 (dotted red), I19254 (dashed magenta), and
QSO1 (dot-dot-dot-dashed green). Filled symbols represent SWIRE/MAMBO
sources with spectroscopic redshifts, open symbols represent sources with
photometric redshifts. Redshift uncertainties are not reported for clarity,
but they are of the order of 0.5 for the photometric redshifts, and
$<$0.1 for the spectroscopic redshits. $f_{1.2\,mm}$ fluxes for the literature sample
are derived from $f_{850\,\mu m}$ when not available. The average 1.2mm/24$\mu$m
flux ratio of 24$\mu$m-selected galaxies~\citep{lutz05} is shown as a dotted
horizontal black line.}
\label{fmam24} 
\end{figure*}

We compare the 1.2\,mm/24\,$\mu$m flux ratio of the SWIRE and literature SMG
samples in Figure~\ref{fmam24}. The SWIRE sources show on average lower flux
ratios than the literature sources at parity of redshift. The difference in
the 1.2\,mm/24$\mu$m can not be simply explained by a redshift difference
but is likely due to an excess in the MIR (i.e. 24$\mu$m) with respect to
the FIR in the SWIRE sample compared to the literature sample. Such an
excess, also visible in the SEDs (see Fig.~\ref{smg_seds_lit_swire}), might
be due to an AGN contribution in the MIR, or to enhanced PAH emission, or to
a deficiency at mm wavelengths due to lower SFRs or higher temperatures.
Indeed objects characterized by warmer dust temperatures are expected to
show lower flux ratios~\citep[e.g.][]{chapman03,lewis05}.

In Figure~\ref{fmam24}, we also show for comparison the 1.2\,mm/24$\mu$m
flux ratios measured in the sample of composite sources from~\citet{lutz05},
and in high-$z$ radio galaxies (HzRG) detected at both 850\,$\mu$m and
24\,$\mu$m~\citep{seymour07}. The 1.2\,mm flux is derived from the
850\,$\mu$m by applying equation~\ref{smm_mm}. These two samples,
from~\citet{lutz05}, and~\citet{seymour07}, show even smaller 1.2\,mm over
24\,$\mu$m flux ratios. In this case, the most plausible explanation for
such small ratios is the contribution from AGN-heated hot dust emission to
the MIR since both samples contain mainly AGNs. Since our sources show
intermediate flux ratios in between those seen in the literature SMGs and
HzRGs, it is quite possible that an AGN contributes to their MIR, more than
typically observed in SMGs and less than observed in HzRGs.

To better investigate the origin of such an excess we consider in more
details the SMGs that fall in the same region covered by the SWIRE sources,
and analyze the MIR spectra when available. From the literature sub-sample
with available spectroscopic redshifts and available 24\,$\mu$m flux (65
sources), we find 3 sources that fall in the same region in the
$f_{1.2mm}/f_{24\mu m}$ $vs$ $z$ diagram as our sources, J123635.59+621424.1
($f_{1.2mm}/f_{24\mu m}$ = 1.35) at $z$ = 2.005 from the GOODS North
field~\citep{borys05}, CSMM\,J163650+4057 ($f_{1.2mm}/f_{24\mu m}$ = 3.07)
at $z$ = 2.384 from the ELAIS-N2 field~\citep{greve04}, and
SMM\,J105238.19+571651.1 ($f_{1.2mm}/f_{24\mu m}$ = 3.34) at $z$ =
1.852~\citep{chapman05}. The first two sources are AGNs, their MIR SEDs are
consistent with a power-law model, their optical spectra show broad emission
lines, and one source is also a luminous X-ray source. The latter source is
instead classified as starburst galaxy based on its optical spectrum, and it
also shows the highest 1.2\,mm/24\,$\mu$m flux ratio of the three selected
sources. This analysis thus reinforces the hypothesis that an AGN component
might be responsible for the lower flux ratios observed in the SWIRE sample,
but this is based on only 2 sources and their SEDs are very different from
those of our sources as a dominating AGN component is clearly seen in the
MIR.

The best and probably unique way to constrain the presence and the
contribution of an AGN component in the MIR is to measure the equivalent
width of the PAH features. A warm dust continuum due to AGN-heated hot dust
will indeed decrease their equivalent width~\citep[see
e.g.][]{clavel00,valiante07,menendez07,sajina07,desai07}. This measurement
can be performed using MIR spectra.

\begin{figure*}
\epsscale{2.0}
\plotone{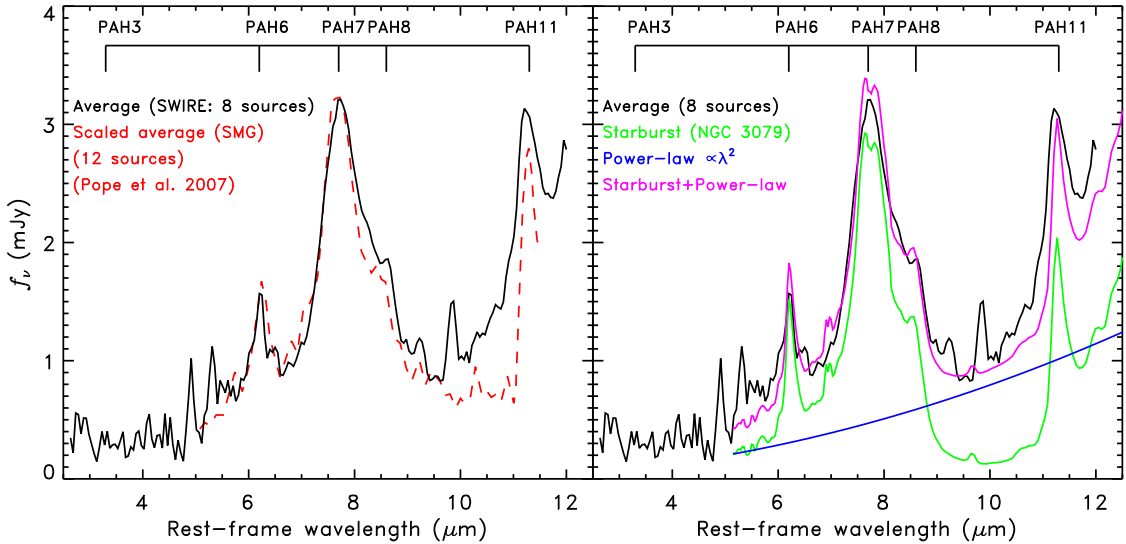}
\caption{Average IRS spectrum of the 8 SWIRE sources with available IRS
spectra (black solid curve). {\it Left panel:} Also shown the composite
spectrum of 12 SMGs from the HDFN~\citep{pope08} after scaling it to match
the SWIRE average spectrum (red dashed curve). {\it Right panel: } IRS
spectrum of a starburst (NGC\,3079; green cure), a power law model with
spectral index 2 representing warm dust continuum emission (blue curve), and
the sum of the two (magenta curve). The expected locations of PAH emission
features are labeled in both panels.}
\label{irs_spe} 
\end{figure*}

Based on the IRS spectra of similar samples and the few available here, a
significant AGN contribution is disfavored for the majority of our sample. 
MIR spectra from IRS are available only for 8 sources in our SWIRE-MAMBO
sample~\citep[][Lonsdale et al., in prep.]{weedman06a}. Altogether, 6
out of these 8 sources show PAH dominated spectra and weak or absent
continuum, implying large equivalent widths and negligible or absent AGN
contribution. They are, in approximative order of PAH strength, LH-07,
LH-06, LH-26, EN2-04, LH-12, and LH-05. The only exceptions are 2 sources
(LH-23/B6, and EN2-03) that show, in addition to PAH emission, a warm dust
continuum that might be associated with an AGN component. We do not
find any trend between the presence of continuum, the strength of the PAHs,
the 1.2\,mm flux, and the $f_{1.2mm}/f_{24\mu m}$ flux ratio. Therefore,
for this analysis we consider the average spectrum of all 8 sources,
independently of their properties. The average IRS spectrum of all 8
SWIRE-MAMBO sources is shown in Figure~\ref{irs_spe}. We also show for
comparison the composite spectrum of 12 SMGs from the HDFN~\citep{pope08},
the IRS spectrum of a starburst, NGC\,3079~\citep{weedman05}, a power-law
model representing emission from warm dust, and the sum of the starburst
spectrum and of the warm dust continuum. The average spectrum of the
SWIRE-MAMBO sources clearly shows PAH features, implying a strong starburst
component, but also a warm dust continuum. The average spectrum is well
reproduced by the sum of the starburst spectrum and the warm dust
continuum.  The continuum is modeled with a power-law model ($f_{\nu}\propto
\lambda^{\alpha_{IR}}$), with spectral index $\alpha_{IR}$=2. We have
applied the same analysis on the average spectrum after removing the 2
sources with larger continuum contamination and for the sources with
$>$2$\sigma$ and $<$2$\sigma$ 1.2\,mm fluxes and the result is still valid
with only small differences that would require a larger sample to better
quantify. The contribution of the power-law component to the average
spectrum is $\sim$34\% in the 6--12\,$\mu$m rest-frame wavelength range.
Assuming that the 24\,$\mu$m flux includes a 34\% contribution from an AGN
on average, the AGN-subtracted $f_{1.2\,\mu m}/f_{24\,\mu m}$ flux ratio
would be 1.5\,$\times$ higher. Even after such a correction, our sources
remain systematically on the lower bound region of the flux ratio
distribution. Note that for consistency we would have to apply a similar
correction to the literature SMGs as well. Indeed, the majority of classical
SMGs also show both PAH features and a weak warm dust continuum in their MIR
spectra~\citep{valiante07,menendez07,pope08}. The PAH equivalenth withs of our
sources and of classical SMGs are indeed very similar, as shown in the left
panel of Figure~\ref{irs_spe}. Thus, the hypothesis that our sources might
have an enhanced 24\,$\mu$m flux due to the contribution from AGN emission,
is disfavored. However, our sources are more luminous at 24\,$\mu$m than
SMGs. Their PAHs and MIR continuum are thus, on average, more luminous than
in SMGs.

\section{Far-infrared properties}\label{fir_seds}

In order to estimate FIR luminosities, and star-formation-rates (SFRs) of
the SWIRE-MAMBO sources, their FIR SEDs need to be characterized. Since
most of the sources are too faint to be detected at 70 and 160\micron\ with
MIPS at the SWIRE sensitivity (typically 18, and 108\,mJy, respectively), we
can only use the observed 1.2\,mm flux and make reasonable assumptions on
their FIR SEDs. We assume that the FIR-mm emission of our sources is
similar to that seen in SMGs, and model their FIR-mm emission with a
greybody model normalized at the observed 1.2\,mm flux or 3$\sigma$ upper
limit in case the flux is $<$2$\sigma$. To constrain the temperature of the
FIR-mm emitting dust, it would be necessary to have at least two flux
measurements at FIR-mm wavelengths. Since the dust temperature can not be
directly estimated for our sources, we fix the temperature to the values
typically observed in SMGs after taking into account the dependency of the
dust temperature on the redshift. From the sample of SMGs with multiple
sub-mm and mm data studied by~\citet{kovacs06}, we derive:
\begin{equation}
T_{dust} = 3.9+9.5\times (1+z) \,\,\,\,\,(K).
\label{temp_z}
\end{equation} 
We use a dust emissivity index $\beta$=1.5 as in~\citet{kovacs06}. Higher
values of $\beta$ give slightly larger temperatures~\citep[see Fig.\,3
in][]{kovacs06}, and larger FIR luminosities. At parity of flux and
temperature, the FIR luminosity derived with $\beta$=2 is on average 2.3
times larger than the FIR luminosity derived with $\beta$=1.5.  Two
sub-mm/mm flux measurements are available only for 4 sources in our sample
(Kovacs et al., in prep.). In these four cases, the estimated temperatures
are consistent with the measured ones, giving support to our assumption. The
dust temperatures range from 23\,K to 41\,K, with a mean value of
33$\pm$4\,K. In order to take into account the redshift uncertainty, the
greybody fits are performed three times: 1) at the redshift of the sources,
$z_{best}$, 2) at a lower redshift,
$z_{-}$=$z_{best}-0.12\times(1+z_{best})$, and 3) at a higher redshift,
$z_{+}$=$z_{best}+0.12\times(1+z_{best})$. The coefficient 0.12 corresponds
to 1.5 times the dispersion found in the comparison between photometric and
spectroscopic redshifts (see \S~\ref{zphot_seds}). The FIR luminosities are
reported in Table~\ref{tab_lum}. The FIR luminosities for the sources with a
$>$2$\sigma$ signal at 1.2\,mm range from 1 to 6.3 $\times$10$^{12}$\,\lsun,
confirming that our SWIRE-MAMBO sources are ULIRGs. This is true for the
majority of the sources also at $z=z_{-}$. A similar luminosity range is
obtained for the rest of the sample, but they are likely upper limits to the
true FIR luminosity.

The greybody models are always below the 70\,$\mu$m, and 160$\,\mu$m upper
limits of each source, implying that the assumed temperatures are never
overestimated. In case of higher temperatures, the FIR luminosities would be
higher than estimated.

We carry out an additional analysis to characterize the average FIR SED of
our sources. We estimate average fluxes at 70\micron\ and 160\micron\ by
co-adding all the 70\micron\ and all the 160\micron\ images to create a
stacked 70\micron\ image and a stacked 160\micron\ image. The co-adding
technique used in this work follows the method described in~\citet{dole06}.
This technique was applied to all MIPS images (24, 70, and 160 \micron) and
only to the sources in the LH field since the number of observed and
detected sources is significantly larger than in the other fields.  The
stacking of the 24\micron\ images was carried out to validate the method by
comparing the stacking result with the mean flux of the single measurements.
In the rest of the analysis, we will use the mean of the single measurements
at 24\micron\ instead of the stacking value. Two sources, LH-18, and LH-28,
were excluded from the stacking because the former is located at the edge of
the 160\,$\mu$m image, and the latter is detected at 70\,$\mu$m. We divided
the remaining 27 sources in the Lockman Hole sample in two sub-groups:
sources with 1.2\,mm fluxes above 2$\sigma$ (12), and below 2$\sigma$ (15).

At each of the positions of the sources in each sub-group, we extract a
72\arcsec$\times$72\arcsec\ sub-image from the full-field SWIRE mosaics. The
sub-images are placed into a stack twice, once with their original
orientation, and again with a 90\deg\ rotation, to remove instrumental
effects such as gradients affecting the stacked flux. However, since the
MIPS maps are already filtered following the techniques described
in~\citet{frayer06}, the rotation should not give a significant difference.
Indeed, a test on the the 70\,$\mu$m stacks with and without images rotation
yields a difference of 1.4\% which is negligible compared to the
uncertainty.  The mean flux image is computed from the stack, and the flux
measured in an aperture (with sky subtraction in an annulus included).  An
uncertainty estimate is derived using a bootstrap technique~\citep{efron93},
in which samples of the subimages are drawn with replacement from the image
stack, and the mean image and aperture flux measurement are recomputed.  The
standard deviation of the aperture flux measurements of 200 repetitions of
this process yields the uncertainty estimate. We have validated these
uncertainty estimates by stacking positions randomly offset by 60 to 80
arcseconds from our source positions; the standard deviation of 500 such
stacks is slightly lower than the uncertainties from the bootstrap method. 
All of the measured stacked fluxes are reported in Table~\ref{tab_coadd}. 
The mean of the single measurements at 24\,$\mu$m agrees in all three cases
with the stacked mean 24\,$\mu$m flux.

The stack of all 27 LH sources results in a detection at both 70 and
160$\mu$m, at $\sim$3.8$\sigma$. The stack of sources $>$2$\sigma$ 1.2\,mm
fluxes is well ($\sim$4$\sigma$) detected at both wavelengths. The stack of
the sample with f$_{1.2mm}<2\sigma$ does not result in a detection at either
wavelength. Note that the significance level of these measurements reflects
mainly differences in noise level due to small number statistics. However,
these measurements provide some constraints on the average FIR flux and
spectrum of our sources.

\begin{figure}
\epsscale{1.0}
 \plotone{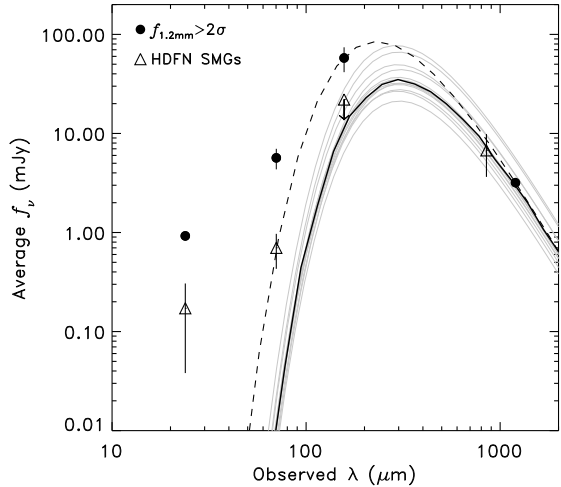}
 \caption{Average FIR SEDs of LH sources with $f_{1.2mm}>$2$\sigma$ (black
 full circles). The 70\,$\mu$m and 160$\mu$m average fluxes correspond to
 the stacked values. The grey curves represent redshifted greybody fits (see
 Eq.~\ref{temp_z}) of all 27 LH sources with $f_{1.2mm}>$2$\sigma$. The
 thick solid black line represents the median of all the greybody fits shown
 in the Figure as grey curves. The dashed black line represents a greybody
 model fit to the average 160\,$\mu$m and 1.2mm fluxes assuming the mean
 redshift, $z$=2.03, and a temperature T=42.5\,K. As expected the median
 1.2\,mm flux of the models and observed are the same, but at 160$\mu$m, the
 median model predicts a flux that is 4.9 times smaller than the measured
 value. The triangles at 24\,$\mu$m and 850\,$\mu$m represent the average
 observed fluxes of the sample of 31 HDFN SMGs~\citep{pope06}, and the
 triangles represent the stacked fluxes at 70\,$\mu$m and 160\,$\mu$m of a
 subsample of 26 sources from the same sample~\citep{huynh07}.}
\label{fir_gb_fits}
\end{figure}

The average FIR SEDs (from 24 to 1200\,$\mu$m, including the stacked 70 and
160\,$\mu$m fluxes) of the LH sub-sample with $f_{1.2\,mm}$$>$2$\sigma$ is
shown in Figure~\ref{fir_gb_fits}. The figure also shows the predicted
average flux at 70 and 160\,$\mu$m assuming the greybody models used to
compute the FIR luminosities. The stacked mean fluxes are well above the
greybody models predictions (solid black curve in Figure~\ref{fir_gb_fits}).
This discrepancy cannot be explain by any of our uncertainty (in redshifts
for example). At 160\,$\mu$m the mean observed flux predicted by the
greybody models is 4 times lower than measured in the stacked images. This
discrepancy suggests that our sources are more FIR luminous and warmer than
estimated by the greybody models that are based on the SMGs temperatures. In
order to fit the SWIRE-MAMBO sources average FIR SED, the dust component
should be on average 10\,K higher than predicted. Indeed, a greybody model
that fits the average FIR SED assuming the mean redshift of the LH
sub-sample, $z$=2.03, has a temperature of 42.5\,K (dashed black curve in
Figure), instead of 32.7\,K as predicted by eq.~\ref{temp_z}. Note that
while the excess at 160\,$\mu$m can be explained by a temperature effect, at
70\,$\mu$m ($\gtrsim$20\,$\mu$m in the rest-frame) a different contribution
from small grains is more likely. In this work, we do not investigate the
difference at 70\,$\mu$m, as we focus our analysis on the FIR part of the
spectrum.

For comparison with the classical SMGs, we consider the FIR properties of
$>$30 SMGs from the HDFN sample~\citep{pope06,pope08}. The mean FIR SED of
this sub-sample is also shown in Figure~\ref{fir_gb_fits}. The 850\,$\mu$m
fluxes of these SMGs range from 1.7 to 20.3\,mJy (corresponding to
0.7--8.3\,mJy at 1.2\,mm), and show a mean value of 5.7\,mJy (corresponding
to 2.4\,mJy at 1.2\,mm), consistent with our SWIRE-MAMBO sources with
$f_{1.2\,mm}$$>$2$\sigma$. Their average 70 and 160\,$\mu$m fluxes, derived
from the stacking of a sub-sample of 26 sources by~\citet{huynh07}, are
significantly lower than those measured in the SWIRE sample. Such a
difference suggests that the HDFN SMGs are, on average, characterized by
cooler dust temperatures and lower FIR luminosities. Note that the
sub-sample used by~\citet{huynh07} does not include 2 sources that are
detected in the MIPS images. However, even including those sources the
average FIR SED of classical SMGs would remain well below that of our
sources.

In summary, the analysis of the average FIR SED of the SWIRE-MAMBO sources
with $f_{1.2mm}$$>$2$\sigma$ indicates that their FIR luminosities are
higher than estimated using the greybody models and that the FIR emitting
dust is at higher temperatures than assumed. On average, our sources are
thus more luminous and warmer than classical SMGs.

\section{Stellar masses}\label{masses}

Our IRAC-based selection corresponds to a rest-frame NIR selection,
therefore, in absence of a significant AGN contribution at these
wavelengths, we are directly sampling the stellar component. Moreover a NIR
selection is minimally affected by dust extinction. Therefore it is possible
to derive accurate stellar masses, by means of spectrophotometric synthesis. 
We focus this analysis on the better observed and larger LH sample.

We use the code Sim-Phot-Spec (SPS), developed in
Padova~\citep{berta04,poggianti01} to estimate the stellar mass in our
sources. The code performs mixed stellar population (MSP)
spectro-photometric synthesis.  The adopted synthesis stellar population
(SSP) library is based on the Padova evolutionary sequences of stellar
models~\citep{fagotto94a,fagotto94b} and isochrones~\citep{bertelli94}, and
was computed by assuming a solar metalicity and a~\citet{salpeter55} initial mass
function (IMF) between 0.15 and 120\,\msun. The SSP spectra are built using
the~\citet{pickles98} stellar spectral library, and extended to the NIR
with the~\citet{kurucz93} stellar atmosphere models. Nebular features are
added based on the ionization code CLOUDY~\citep{ferland96}. The spectra
thus obtained provide a reliable description of simple stellar generations
up to $\sim$3\,$\mu$m in the rest-frame. Note that our models do not take
fully into account the behavior of AGB stars, in other words they
underestimate the contribution from intermediate-age SSPs. We estimate that
this results in overestimating stellar masses of at most a factor of
2~\citep[see a discussion in][]{berta07b}. Another overestimating factor
comes from the assumption of a Salpeter IMF, indeed a~\citet{kroupa01} or
a~\citet{chabrier03} IMF would yield about 50\% lower
masses~\citep[see][]{berta07b}.

The mixed stellar population synthesis combines several phases, adopting a
different star formation rate (SFR) for each age.  This code is well suited
for modeling stellar populations of star-forming galaxies, because it allows
several, independent episodes of star formation during the life of a galaxy,
for example resembling multiple merger-driven starburst events.
Age-selective extinction is applied, assuming that the oldest stars have
abandoned the dusty medium long ago. Since disk populations are on average
affected by a moderate extinction~\citep[\av$\sim$$<$1 mag,
e.g.][]{kennicutt98}, we adopt a maximum allowed absorption for stars older
than 1 Gyr of \av\ = 0.3 -- 1.0 magnitudes. For younger populations the
color excess gradually increases, but is limited to \av$\leq$5.  The best
fit is sought by means of $\chi^2$ minimization.  In case of non-detections
3$\sigma$ upper limits are adopted. For each source in the sample, the code
explores the SFR-extinction parameter space using an Adaptive Simulated
Annealing algorithm and records all the attempted models. The uncertainty in
the stellar mass estimate is then given by all those solutions whose
$\chi^2$ values lie within 3$\sigma$ from the best fit. Details on the
fitting method are described in~\citet{berta07b}.

Assuming that the total energy absorbed in the ultraviolet (UV)-optical domain is
processed by dust in the thick molecular clouds embedding young stars and
re-emitted in the MIR and FIR (5--1000\,$\mu$m), we exploit the 24\,$\mu$m
observed flux to constrain the amount of dust and the luminosity of young
stellar populations~\citep[see][ for more details]{berta04}.

\begin{figure*}
\epsscale{2.0}
\plotone{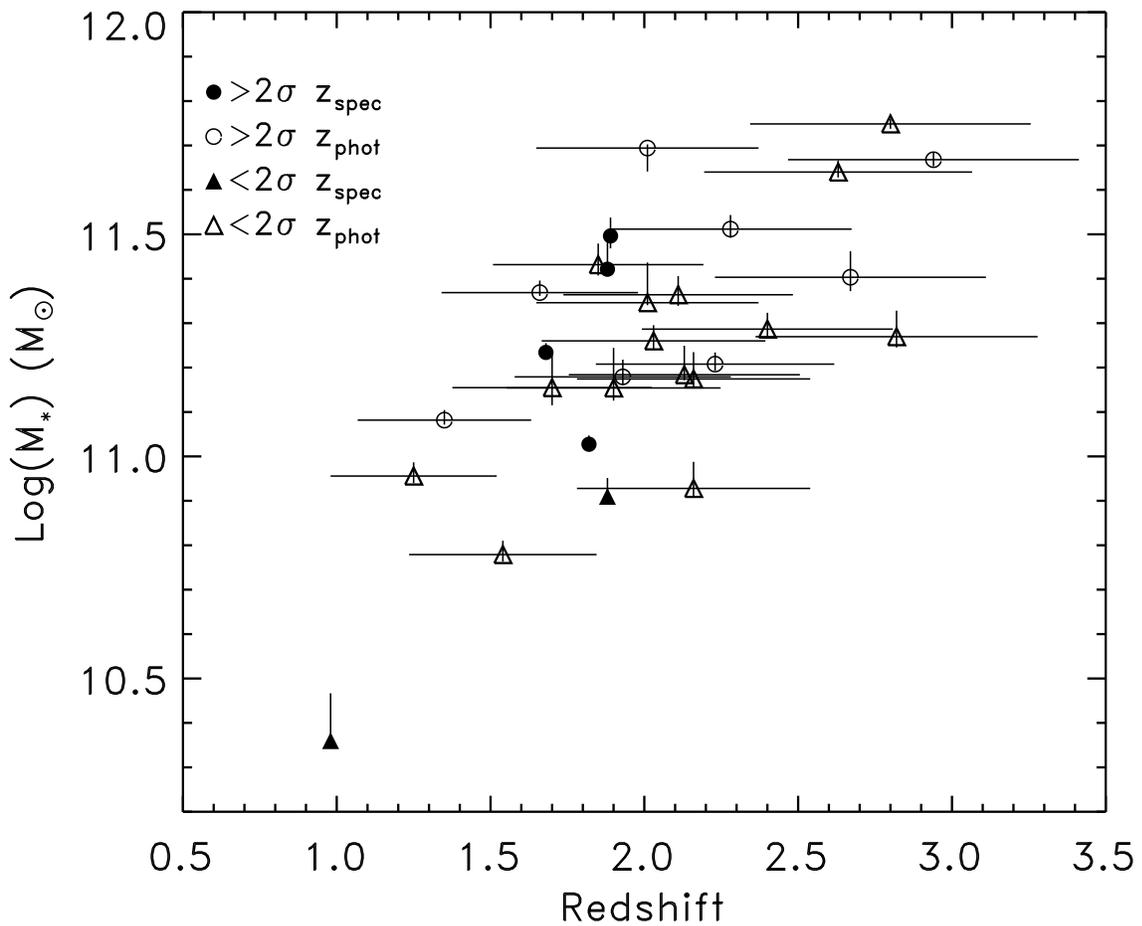}
\caption{Stellar masses as a function of photometric (open symbols) or
spectroscopic (full symbols) redshifts for the SWIRE Lockman Hole sample (29
sources).  Circles represent sources with $f_{1.2\,mm}$$>$2$\sigma$ and
triangles represent sources with $f_{1.2\,mm}$$<$2$\sigma$.  Uncertainties
on the stellar masses correspond to 2$\sigma$.}
\label{ol_mass_z}
\end{figure*}

The estimated stellar masses and associated uncertainties are listed in
Table~\ref{tab_masses}. The stellar masses range from 2.3 to 56 $\times
10^{10}$\,\msun, with a median value of 1.8$\times 10^{11}$\,\msun. 
Although the near-IR SED is dominated by stellar emission, it is possible
that an AGN component is also present. Thus these estimates should be
considered as upper limits to the true stellar masses. It is also possible
that the derived stellar masses overestimate the true masses because our
models do not take into account the TP-AGB contribution~\citep{maraston05}.
Indeed such a contribution might yield lower masses, especially at the
redshifts of our sources. The stellar masses as a function of redshift are
shown in Figure~\ref{ol_mass_z}. 
We find that at parity of redshift, the stellar masses can vary by a factor
of 4. No difference in stellar mass is observed between 1.2\,mm detected and
non-detected objects. Similar masses are observed in objects with a wide
range of luminosities and viceversa. For example, all high-$z$ ($z>$1.5)
sources with available spectroscopic redshifts have similar FIR luminosities,
$\sim 10^{12.3}$\lsun, but their stellar masses can vary by a factor of 5,
i.e. M$_*$=8--20$\times 10^{10}$\msun. 

\subsection{NIR luminosities}

\begin{figure*}
\epsscale{2.0}
\plotone{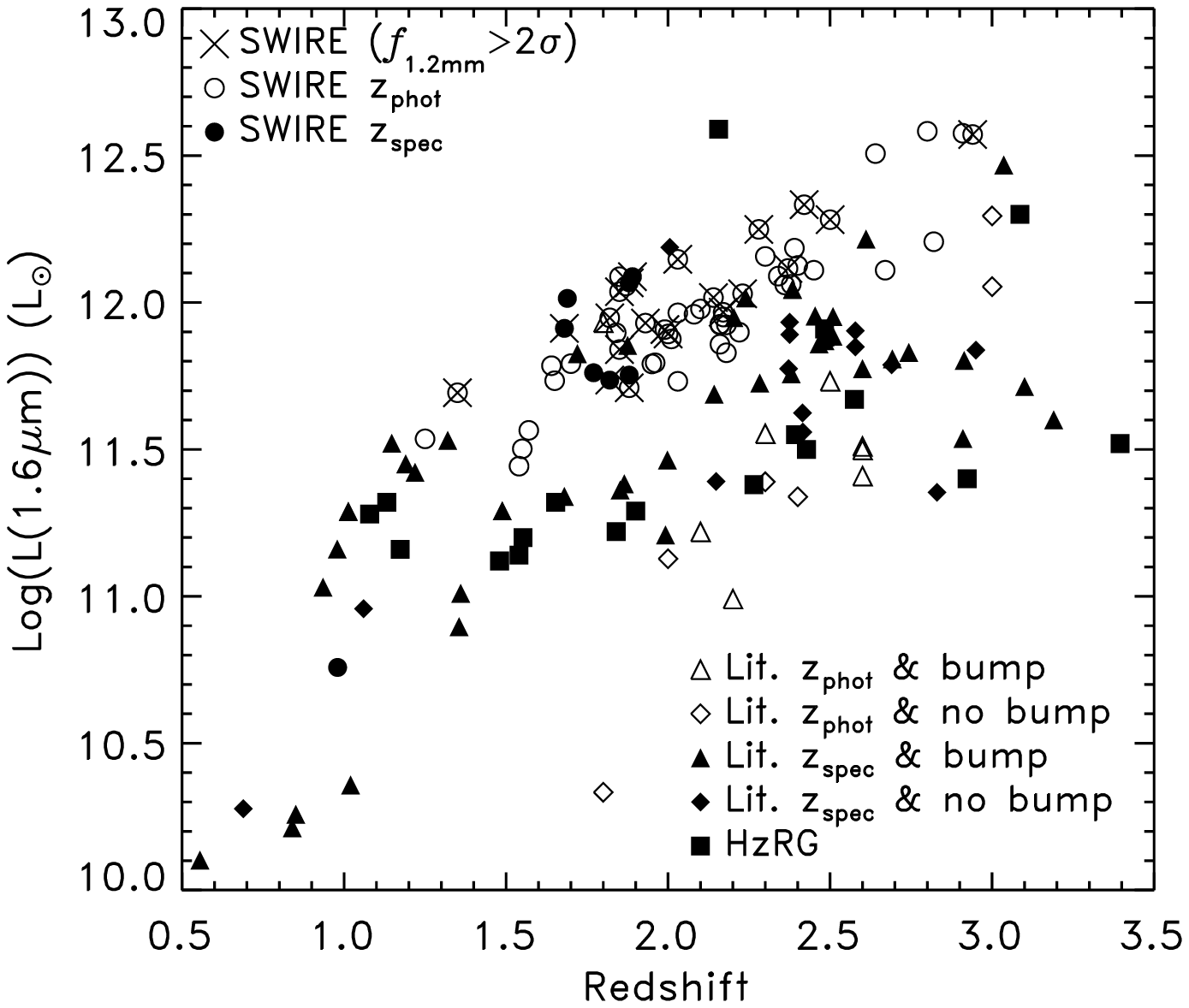}
\caption{Monochromatic luminosity at 1.6\,$\mu$m in the
rest-frame for SWIRE/MAMBO sources (circles), for literature SMGs (triangles
and diamonds), and for high-$z$ radio galaxies (squares). Full symbols
represent sources with spectroscopic redshifts and open symbols with
photometric redshifts. Triangles represent sources with a typical stellar
bump in the IRAC bands and diamonds sources where the bump is not present or
not detected. Large crosses are overplotted on SWIRE/MAMBO sources with
$f_{1.2\,mm}$$>$2$\sigma$.}
\label{lnir_z}
\end{figure*}

In order to carry out a consistent comparison between the stellar masses of
the SWIRE-MAMBO sources with the SMGs found in the literature, we compare
their rest-frame luminosity at 1.6\,$\mu$m, where the stellar emission
peaks. The rest-frame luminosity at 1.6\,$\mu$m is estimated by
interpolating the observed IRAC fluxes. The estimated values as a function
of redshift are reported in Figure~\ref{lnir_z}. They
range from 5.8$\times$10$^{10}$\,\lsun\ to 3.8$\times$10$^{12}$\,\lsun, and
have a median value of 9.5$\times$10$^{11}$\,\lsun. We compare the
luminosities of sources with a spectroscopic redshift and those with only a
photometric redshift, and sources with 1.2\,mm fluxes greater or lower than
2$\sigma$. No apparent difference in NIR luminosity is observed among these
sub-samples. We also divide the SMGs from the literature in two sub-groups
(shown as triangles and diamonds in Figure~\ref{lnir_z}), those with a
clear stellar bump in the IRAC bands and those for which a bump is not
visible because of a different spectral shape or because there are no
detections at $\lambda$=5.8 and 8.0\,$\mu$m. There is no significant
difference in NIR luminosities between the SMGs with and without a stellar
bump. The SWIRE-MAMBO sources are systematically more luminous in the NIR,
suggesting that they might be more massive than literature SMGs. On average
our sources are 3 times more luminous at 1.6\,$\mu$m than sub-mm selected
SMGs. This result is not surprising since our selection is based on a
detection at 5.8$\mu$m where the SWIRE 5$\sigma$ limit is
$\simeq$50\,$\mu$Jy, while the literature sample is, on average,
characterized by lower 5.8\,$\mu$m fluxes, and both samples are at similar
redshifts.

\section{AGN activity}\label{agn_contr}

It is well established that starburst sources dominate the bulk
population of SMGs with only a minor contribution from AGN. Indeed, although
a weak AGN emission is observed in a significant fraction of SMGs, at X-ray
energies~\citep{alexander05}, in optical spectra lines~\citep{chapman05},
and in MIR spectra~\citep{valiante07,menendez07,pope08}, its contribution to their
bolometric luminosity is usually modest ($<$10--20\%).

In \S~\ref{colors}, the contribution from an obscured AGN to the observed
MIR emission of our sources was invoked to explain their unusual colors
and SEDs. Based on the available IRS spectra, we estimate, on average, a
34\% contribution from a warm dust continuum in the MIR (6--12\,$\mu$m in
the rest-frame). Such a component might be generated by an AGN, or by HII
regions~\citep{laurent00,dopita06}. However, this estimate is based on only
8 sources. Moreover, the IRS targets include the brightest 24\,$\mu$m
sources of the sample and might thus be biased towards sources with enhanced
PAH emission or larger AGN contribution~\citep{brand06}. To further
investigate the presence of an AGN in our sample, we searched for possible
signatures using X-ray and radio data.

X-ray observations from \chandra\ of moderate depth (70\,ks) are available
for 3 LH sources, LH-10, LH-19, and LH-28, from the \chandra/SWIRE
survey~\citep{polletta06}, and the CLASX survey~\citep{yang04}. Although the
XMM field has a wide shallow (10\,ks) X-ray coverage from \xmm\, only one
source, XMM-03, has been observed by \xmm~\citep{pierre07}, and it is at the
edge of the field. Out of 4 sources with X-ray coverage, only LH-10 is
detected in the X-ray. The X-ray 0.4--8 keV flux of LH-10 is
2.1$\times$10$^{-15}$\,\ergcm2s, and the hardness ratio is
HR=$-$0.7~\citep{yang04}, as expected for a steep unabsorbed power-law
spectrum. The estimated 0.4--8 keV luminosity is 5.7$\times 10^{43}$\,\ergs,
consistent with being AGN powered and with those measured in X-ray detected
SMGs in HDFN~\citep{alexander05b}. Although, based on the hardness ratio,
the X-ray spectrum appears unabsorbed, the derived L$_{MIR}$/L$_{X-ray}$
ratio, $\simeq$100, is much higher than commonly observed in AGNs,
0.3--10,~\citep[see e.g.][]{franceschini05,polletta07}. The large
L$_{MIR}$/L$_{X-ray}$ ratio suggests that the X-ray luminosity is
underestimated and that the observed X-ray emission corresponds to a small
fraction of the intrinsic X-ray emission, perhaps due to a reflected
component (warm scattering), while the intrinsic X-ray emission is
completely absorbed. The lack of X-ray detections of the other 3 sources
does not rule out the presence of an absorbed X-ray luminous AGN
($L_{X-ray}>10^{44}$\,\ergs) or an unabsorbed AGN of moderate luminosity
($L_{X-ray}<10^{44}$\,\ergs). Indeed the detection limit of the \chandra\
observations corresponds to X-ray luminosities $\geq$10$^{44}$\,\ergs\ for
an unabsorbed AGN at the redshift of our sources. An AGN of lower luminosity
or absorbed, as typically found in SMGs, would not be detected at our
depths.

Deep VLA radio observations at 20\,cm are available in a 0.4 deg$^2$ area
centered on the \chandra/SWIRE field~\citep[][ Owen et al., in
prep.]{polletta06} in the LH, and shallow radio observations are available
from the VLA FIRST survey in all fields\footnote{The VLA FIRST source catalog and images
are available from http://sundog.stsci.edu.}~\citep{condon98}. LH-11, and
LH-22 are the only sources that fall in the deep VLA field and are both
detected, with 20\,cm fluxes of 82 and 51\,$\mu$Jy, respectively. A third
source, LH-28, is detected by the FIRST survey. The radio fluxes are
reported in Table~\ref{tab_mm}. The radio source associated with LH-28 is
extended, 6\farcs66$\times$3\farcs16, corresponding to
57\,kpc$\times$27\,kpc. The bright radio flux and radio extension, combined
with the lack of mm detection, indicate that the radio emission of LH-28 is
powered by a jet and it is, thus, an AGN. LH-11, and LH-22 are not resolved
in the radio images, $<$1\farcs24 and 0\farcs87, respectively. Although
their 1.4\,GHz rest-frame radio luminosities are large, respectively,
10$^{28.2}$ and 10$^{27.8}$ W Hz$^{-1}$, they are not in excess of what is
expected based on the FIR-radio relationship found in both starburst and
radio-quiet AGNs~\citep{condon98}. Therefore, the available radio data do
not imply or rule out the presence of an AGN in these two luminous radio
sources.

In summary, the available X-ray and radio data indicate that at least 1 out
of 3 sources contains an AGN. This fraction is consistent with that derived
from IRS spectroscopy ($\geq$30\%), but all these samples are limited by
small number statistics and there is no overlap among them, leaving the
question on the AGN contribution and role in our SWIRE-MAMBO sample
unanswered. Moreover, the subset of sources observed with the IRS might be
biased towards a certain type of sources, for example with enhanced PAHs,
and might not be representative of the whole SWIRE-MAMBO sample. Yet, since
there is no apparent difference in the 1.2\,mm/24\,$\mu$m flux ratios of the
X-ray and radio AGNs and the IRS sample, the estimated AGN contribution
might be valid for the whole SWIRE-MAMBO sample. However, altough it is
quite likely that an AGN is present in a large fraction of our sources, it
is not expected to significantly contribute to their MIR emission, as it is
the case for the classical SMGs~\citep[see e.g.][]{pope08}. Indeed, even by
taking this contribution into account, we still find smaller $f_{1.2\,\mu
m}/f_{24\,\mu m}$ flux ratios, larger FIR luminosities, and a FIR peak at
shorter wavelengths than in classical SMGs. Thus, as already stated in
\S~\ref{colors}, it is unlikely that the difference between the SWIRE-MAMBO
sources and classical SMGs is due to a more important AGN contribution in
our sample.

\section{Discussion}\label{discussion}

In the previous sections we have investigated the multi-wavelength
properties of the SWIRE-MAMBO sources selected here and compared them with
those of classical SMGs. Our sources are characterized by redshifts,
1$<z<$3, and mm fluxes in the same range as classical SMGs. Compared to
classical SMGs, they are in general brighter at mid and near infrared
wavelengths. Their FIR luminosities, constrained for the sources detected at
1.2\,mm, are in the ULIRG range, $\geq$10$^{12}$\,\lsun. Their FIR emission
is likely dominated by warmer dust than estimated in classical SMGs, but
this result needs to be confirmed by additional sub-mm data. Their stellar
masses are large, 10$^{10.3-11.8}$\,\msun, and their NIR luminosities and,
likely stellar masses, are also higher by about a factor 3 than in classical
SMGs. Although some of these estimates are characterized by large
uncertainties, it is a clear that our sources do show some differences
compared to classical SMGs, or that they belong to a subclass of SMGs with
larger MIR/FIR flux ratios than the bulk of SMGs (Figure~\ref{fmam24}).
Indeed, SMGs with a relatively small 1.2\,mm/24\,$\mu$m flux ratio
represents only a minority of all SMGs reported in the literature.

The possible origin of the observed differences between our sources and
classical SMGs are discussed here. Warm dust emission is an indicator of
strong overall heating of the dust. A large infrared luminosity and faint
optical/UV emission implies elevated dust opacity, which in turn might imply
a larger density of interstellar dust closer to the heating sources, or a
more efficient heating mechanism relative to the
dust~\citep{dale07,kovacs06}. Dust heating is predominantly fueled by the
emission from massive ($>$8\,\msun) stars in star-forming
regions~\citep{kovacs06}, or by an AGN radiation field. The peak of IR
emission of cold dust can thus shift towards shorter wavelengths in case (1)
the radiation field gets higher because of younger stellar
populations~\citep{piovan06a}, or (2) a significant contribution from
AGN-heated dust is present. Thus, the difference in the FIR emission peak
between our sample and classical SMGs (see \S~\ref{fir_seds}) might be an
indication of more massive stars relative to dust mass, or of a more
significant AGN contribution in our SWIRE-MAMBO sources compared to
classical SMGs.

The latter scenario has been investigated by measuring the intensity
and equivalent width of the PAH features. Indeed, weak PAH emission are
indicative of an important AGN contribution.  Since, the average MIR
spectrum, and thus the PAH features equivalent width, of our sources and of
classical SMGs are consistent (see Figure~\ref{irs_spe}), the idea of having
a significant AGN contribution is disfavored. This result supports the
alternative scenario of having a higher contribution of massive stars
relative to the dust mass. However, the available data do not allow us to
test this hypothesis and rule out other explanations, perhaps related to the
star forming regions geometry and distribution~\citep[see e.g.][]{farrah08}.

Our results indicate that our selection has revealed a class of ULIRGs
that is under-represented in current SMG samples. The SWIRE-MAMBO sources
indeed cover a different region in the luminosity-temperature ($L-T_d$)
diagram than classical SMGs and bridge the gap between the latter and local
ULIRGs~\citep[see][]{chapman03,lewis05}. Similar results have been obtained
by a study on a similar sample of \spitzer-selected sources with known
spectroscopic redshifts from strong PAH features, and multiple FIR
detections (Younger et al., in prep.).

\section{Summary and conclusion}\label{summary}

Using the SWIRE survey and the MAMBO camera on the IRAM 30m telescope, we have
identified 21 ULIRGs at $z\simeq$1--3 with average 1.2\,mm fluxes of
2.9$\pm$0.2\,mJy. They are characterized by starburst-dominated SEDs and by
unusually large MIR/FIR flux ratios. These sources were discovered among a
sample of 61 ULIRG $z>$1 candidates selected in the SWIRE survey that were
followed-up with MAMBO observations at 1.2\,mm.

The primary selection criteria of the SWIRE sample of ULIRG candidates are a
peak in the IRAC 5.8$\mu$m band due to the red-shifted near-infrared
spectrum of evolved stars, a bright ($>$400\,$\mu$Jy) detection at 24$\mu$m,
and a faint optical counterpart (\rp$>$23). The average 1.2\,mm flux of the
whole sample is 1.5$\pm$0.2\,mJy.

The optical-MIR SEDs of the selected sources are consistent with those of
starburst galaxies. Photometric redshifts are estimated by fitting the SEDs
with galaxy templates. For eight sources, spectroscopic redshifts from MIR
spectra obtained with IRS on board of \spitzer\ are available and are in
reasonable agreement with the estimated photometric redshifts.

Our analysis focuses on 29 sources in the Lockman Hole field where the
average 1.2\,mm flux (1.9$\pm$0.3\,mJy) is higher than in the XMM-LSS,
ELAIS-N1 \& ELAIS-N2 fields (1.1$\pm$0.2\,mJy). However, we also present
basic results for 32 sources from the other fields, which were less
thoroughly observed. The fraction of LH sources with 1.2\,mm fluxes
$\geq$2$\sigma$ is $\sim$40\% (24\% at 4$\sigma$; see Table~\ref{tab_stat}).
This fraction is much higher than previously found for \spitzer\ MIR
selected samples~\citep[23\%, 18\%, and 10\% at $\geq$2, 3, and 4$\sigma$,
respectively; ][]{lutz05}. We attribute this difference to the fact that
earlier samples favored AGN-dominated, rather than
starburst-dominated systems. Our sample, on the other hand, shows
systematically lower 1.2mm/24$\mu$m flux ratios than the majority of
\spitzer-detected sub-millimeter-selected SMGs. Such a difference is
attributed to a bias in our selection in favor of 24$\mu$m-bright systems
compared to the selection of SMGs.

Far-infrared luminosities, and star formation rates are estimated assuming a
$z$-dependent dust temperature as observed in classical SMGs, and
normalizing a greybody model to the observed mm flux.  Average 70 and
160$\mu$m fluxes from stacked \spitzer/MIPS images are measured. The mean
FIR stacked fluxes are $\sim$4 times larger than the fluxes predicted by the
greybody models, suggesting higher FIR luminosities and a FIR peak at
shorter wavelengths than in classical SMGs. The SWIRE-MAMBO sources are
ULIRGs in the $z$=1.5--3 range with FIR luminosities from 10$^{12}$ to
10$^{13.3}$\,\lsun. Stellar masses, estimated by modeling the
optical-infrared data with stellar synthesis populations, are also large,
$\sim$0.2--6$\times$10$^{11}$\msun, suggesting that these systems might be
the precursors of the most massive ellipticals.

Compared to sub-millimeter selected galaxies (SMGs), the SWIRE-MAMBO sources
are among those with the largest 24$\mu$m/millimeter flux ratios. The origin
of such large ratios is investigated by comparing the average mid-infrared
spectra and the stacked far-infrared spectral energy distributions of our
sources and classical SMGs. 

The mid-infrared spectra, available for a handful of sources, show strong
PAH features and a warm dust continuum. The warm dust continuum contributes
to $\sim$34\% of the mid-infrared emission, and is likely associated with an AGN
component. This constribution is consistent with what is found in SMGs, and
indeed the PAH equivalent widths are, on average, similar in our sources and
in SMGs. Thus, even taking into account the AGN component, the
24$\mu$m/1.2\,mm flux ratios remain larger than typically observed in SMGs.
The hypothesis of contribution from AGN emission to the MIR is thus
disfavored, and it is more likely that the large ratios are due to more
luminous PAH emission that in classical SMGs.

The analysis of the stacked far-infrared SEDs yields warmer dust
temperatures than typically observed in classical SMGs. Thus, our selection
has favored a warm class of ultra-luminous infrared sources at
high-$z$ that is rarely found in current SMG samples. Indeed classical SMGs
are not common among bright 24\,$\mu$m sources as the SWIRE-MAMBO sample
(e.g. only about 20\% of SMGs have $f_{24\mu m}>$0.4\,mJy). Our sample
occupies a region in the $L-T_d$ diagram that is under-populated by
classical SMGs, implying that ULIRGs at high-$z$ might exhibit a similar
range in dust temperatures as local ULIRGs~\citep[see][]{chapman03,lewis05}.
The analysis of our sample suggests that warmer ULIRGs sre brighter MIR
sources and might exhibit more luminous PAH emission than cooler ULIRGs.
However, to confirm such a correlation, an analysis of the
MIR spectra for a larger sample of 24\,$\mu$m-selected objects with known mm
fluxes would be necessary. This analysis will be soon possible with other
\spitzer\ samples that have been or will be observed with the IRS and with
MAMBO (Younger et al., in prep., Lagache et al., in prep.).

Our sample is the largest \spitzer-selected sample detected
at millimeter wavelengths currently available.

\acknowledgments

We thank the referee for useful comments that improved the paper. CJL and MP
thank Bruce Elmegreen and Jakob Walcher for helpful discussion. We are
grateful to Elisabetta Valiante for providing the average IRS spectrum of a
sample of SMGs. MP acknowledges financial support from the Marie-Curie
Fellowship grant MEIF-CT-2007-042111. This work is based on observations
made with the IRAM 30m Telescope and the {\it Spitzer Space Telescope}. The
IRAM 30m Telescope is funded by the Centre National de la Recherche
Scientifique (France), the Max-Planck Gesellschaft (Germany), and the
Instituto Geografico Nacional (Spain). The {\it Spitzer Space Telescope} is
operated by the Jet Propulsion Laboratory, California Institute of
Technology under NASA contract 1407. Support for this work, part of the {\it
Spitzer Space Telescope} Legacy Science Program, was provided by NASA
through an award issued by the Jet Propulsion Laboratory, California
Institute of Technology under NASA contract 1407. We thank the IRAM staff
for their support during the observations, and E. Kreysa and his group for
providing the MAMBO bolometer array. This research has made use of the
NASA/IPAC Extragalactic Database (NED) which is operated by the Jet
Propulsion Laboratory, California Institute of Technology, under contract
with the National Aeronautics and Space Administration.

Facilities: \facility{\spitzer(IRAC,MIPS)}, \facility{IRAM(MAMBO)}.

\appendix
\section{SED fits and photometric redshifts}\label{app}

The best and acceptable fits to the optical ($Ugriz$) and MIR
(3.6--24\,$\mu$m) SEDs of each source, obtained as described in
\S~\ref{zphot_seds} are shown in Figures~\ref{sed_fits_zphot_ol},
\ref{sed_fits_zphot_o1}, \ref{sed_fits_zphot_o2} and
\ref{sed_fits_zphot_ox}. We report a maximum of three solutions, those with
the lowest 3 $\chi^2$. In case a spectroscopic redshift is available, the
best-fit template at the spectroscopic redshift is reported as dotted curve.
Although the templates extend to FIR wavelengths, the FIR data are not used
in the fitting procedure and in the $\chi^2$ computation. The parameters
(template, and redshift) of each solution are reported in
Table~\ref{tab_fits}.

Eight sources have spectroscopic redshifts and are shown in
Figures~\ref{sed_fits_zphot_ol} and \ref{sed_fits_zphot_o2}.  It can be
appreciated from this figure that with only the broad NIR bump to constrain
the photometric redshift, using the broad IRAC filters, a redshift accuracy
of 10\%\ in (1+$z$) is a reasonable expectation. In some cases, a NIR bump
can be present in sources at lower redshift. Indeed the presence of warm
dust continuum can make the NIR bump or the turnover at 8$\mu$m less
pronounced and even shift the stellar peak to longer
wavelengths~\citep{berta07,daddi07}. Also strong emission from the 3.3$\mu$m
PAH feature falling into the 5.8\,$\mu$m filter can produce a peak in the
IRAC SED that would be wrongly interpreted if no optical or NIR data are
available. These two effects increase the uncertainty associated with the
photometric redshift estimate. Some examples can be seen in the fit to
source LH-12 ($z_{spec}$=1.89 and $z_{phot}$=2.65) and LH-23
($z_{spec}$=0.98 and $z_{phot}$=1.21) in Figure~\ref{sed_fits_zphot_ol}.
\begin{figure*}
\epsscale{1.0}
 \plotone{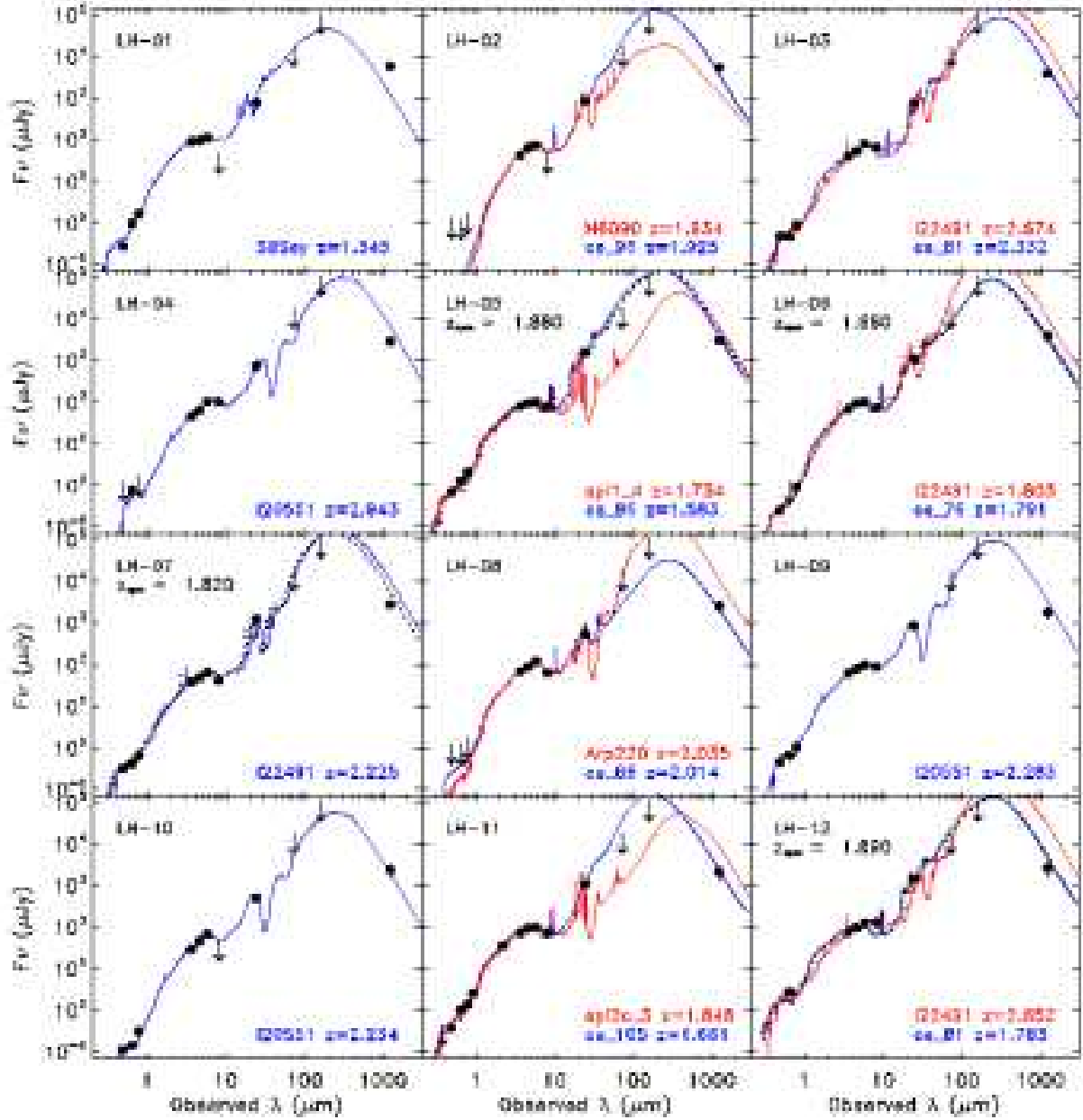}
 \caption{Optical-IR SEDs of MAMBO sources in the Lockman Hole field. The
 solid curves represent different fits to the optical-MIR data (up to
 24\,$\mu$m). Blue, red and green curves corrrespond to fits with increasing
 $\chi^2$.  The template name and photometric redshifts of each fit are
 annotated. The dotted curve corresponds to the best-fit template plotted at
 the spectroscopic redshifts, The source ID number, and the spectroscopic
 redshift when available, are reported on the upper left corner in each
 panel. Downward arrows represent 5$\sigma$ upper limits at optical and
 infrared wavelengths, and 3$\sigma$ upper limits at 1.2\,mm.}
\label{sed_fits_zphot_ol}
\end{figure*}
\addtocounter{figure}{-1}
\begin{figure*}
 \plotone{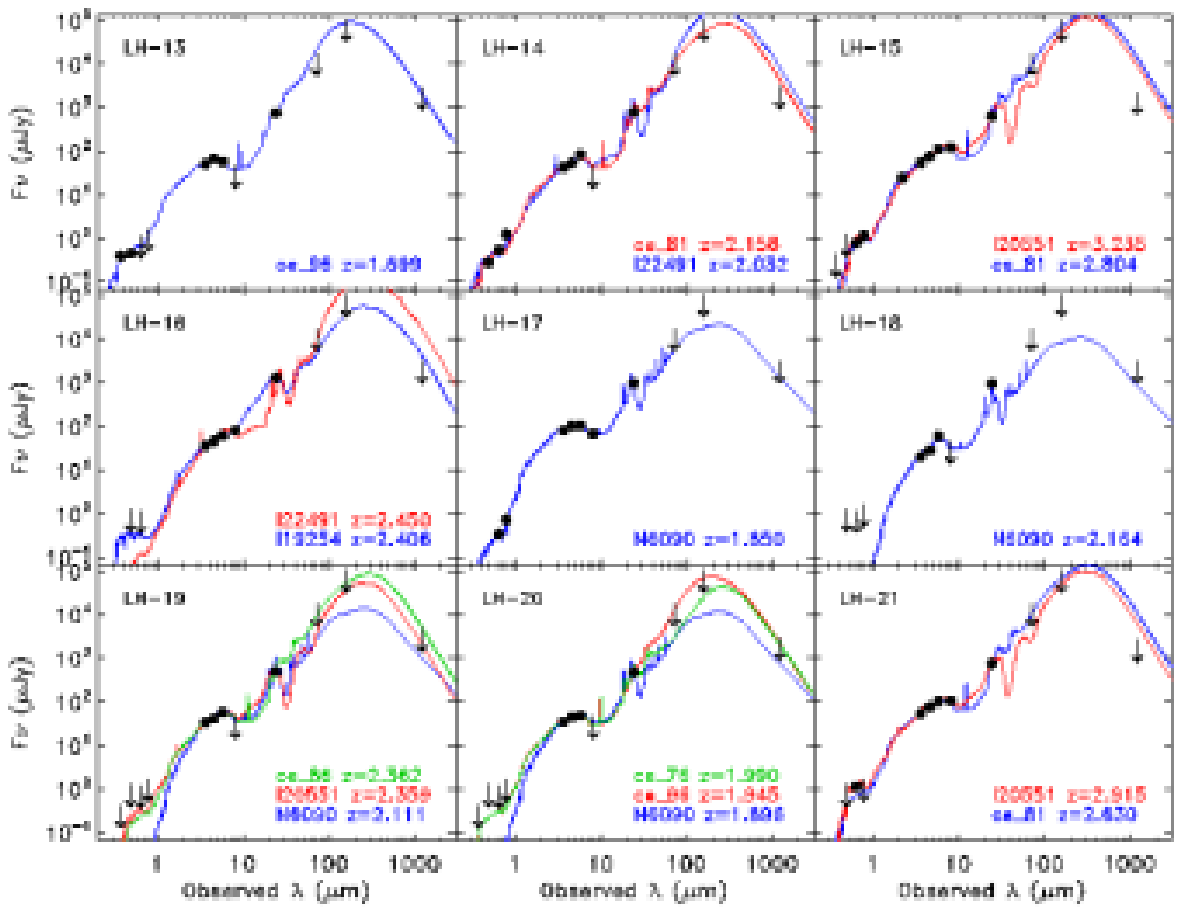}
 \caption{\it Continued}
\end{figure*}
\addtocounter{figure}{-1}
\begin{figure*}
 \plotone{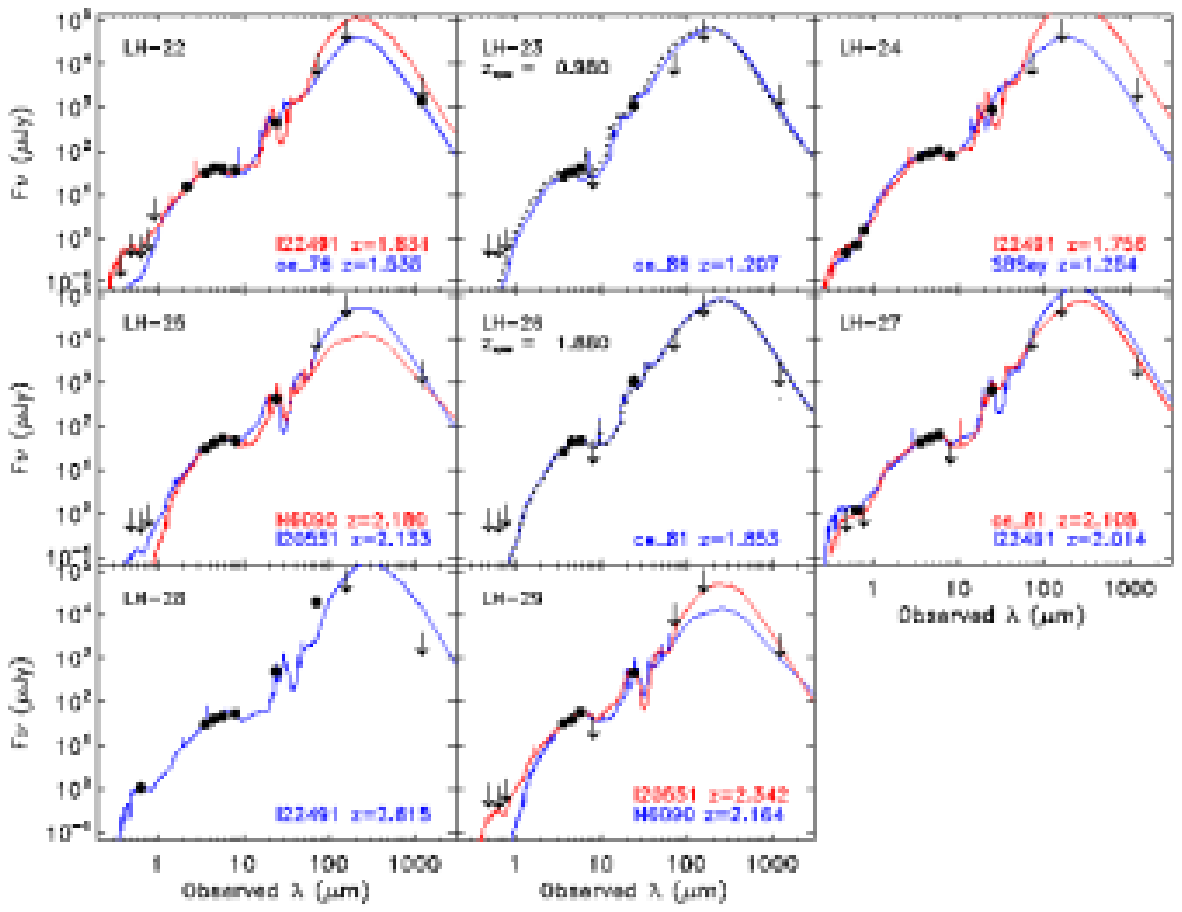}
 \caption{\it Continued}
\end{figure*}

\begin{figure*}
\epsscale{1.0}
 \plotone{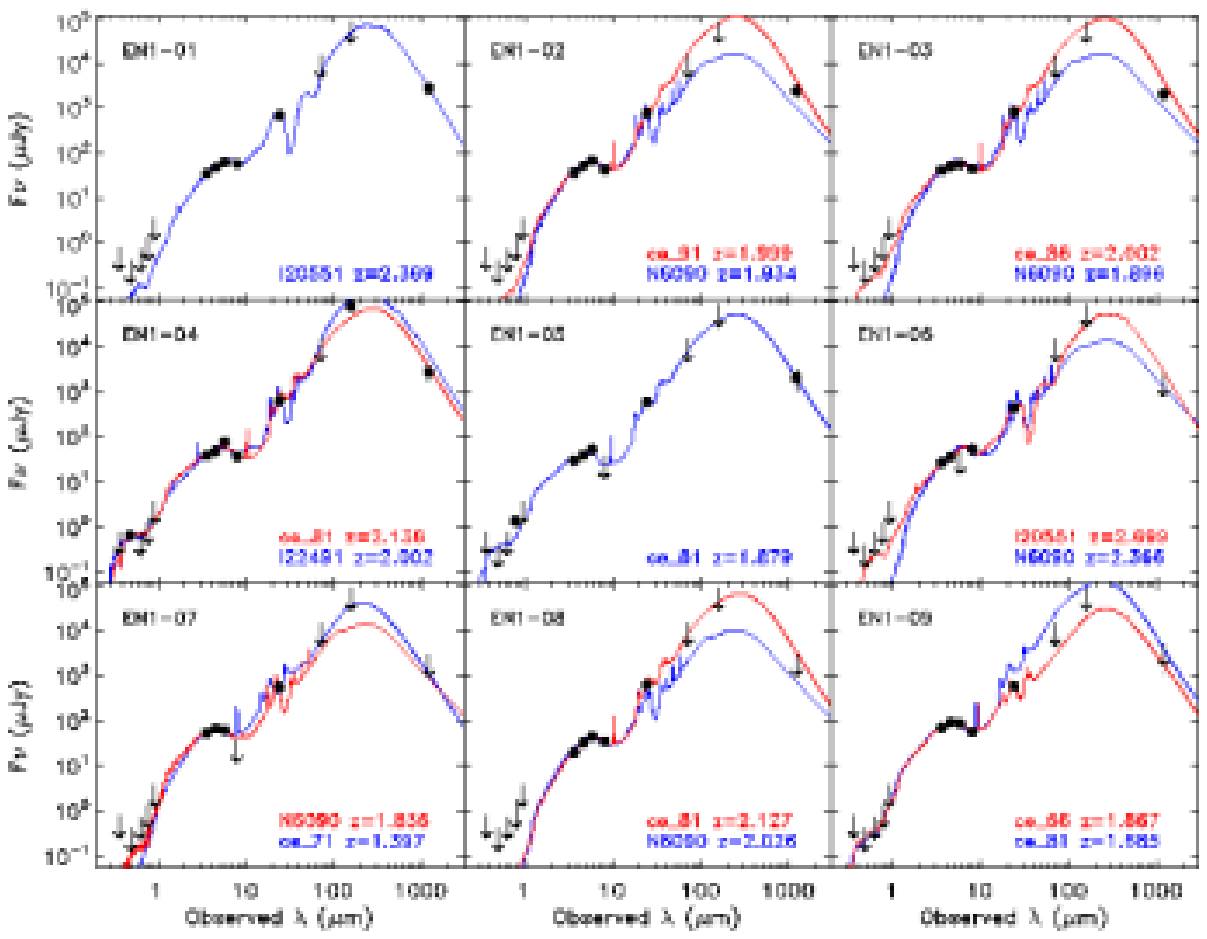}
 \caption{Optical-IR SEDs of MAMBO sources in the ELAIS-N1 field.
  Symbols as in Figure~\ref{sed_fits_zphot_ol}.}
 \label{sed_fits_zphot_o1}
\end{figure*}
\addtocounter{figure}{-1}
\begin{figure*}
 \plotone{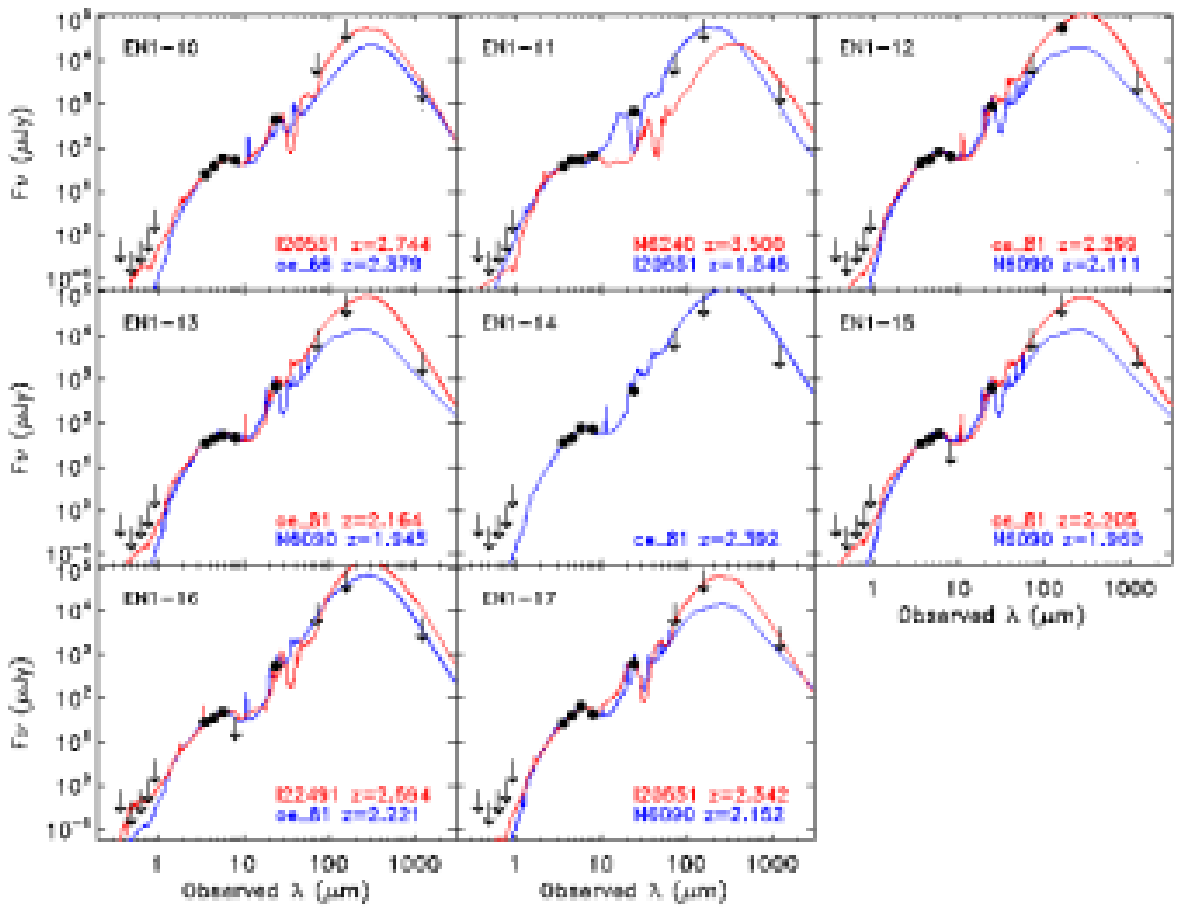}
 \caption{\it Continued}
\end{figure*}

\begin{figure*}
\epsscale{1.0}
 \plotone{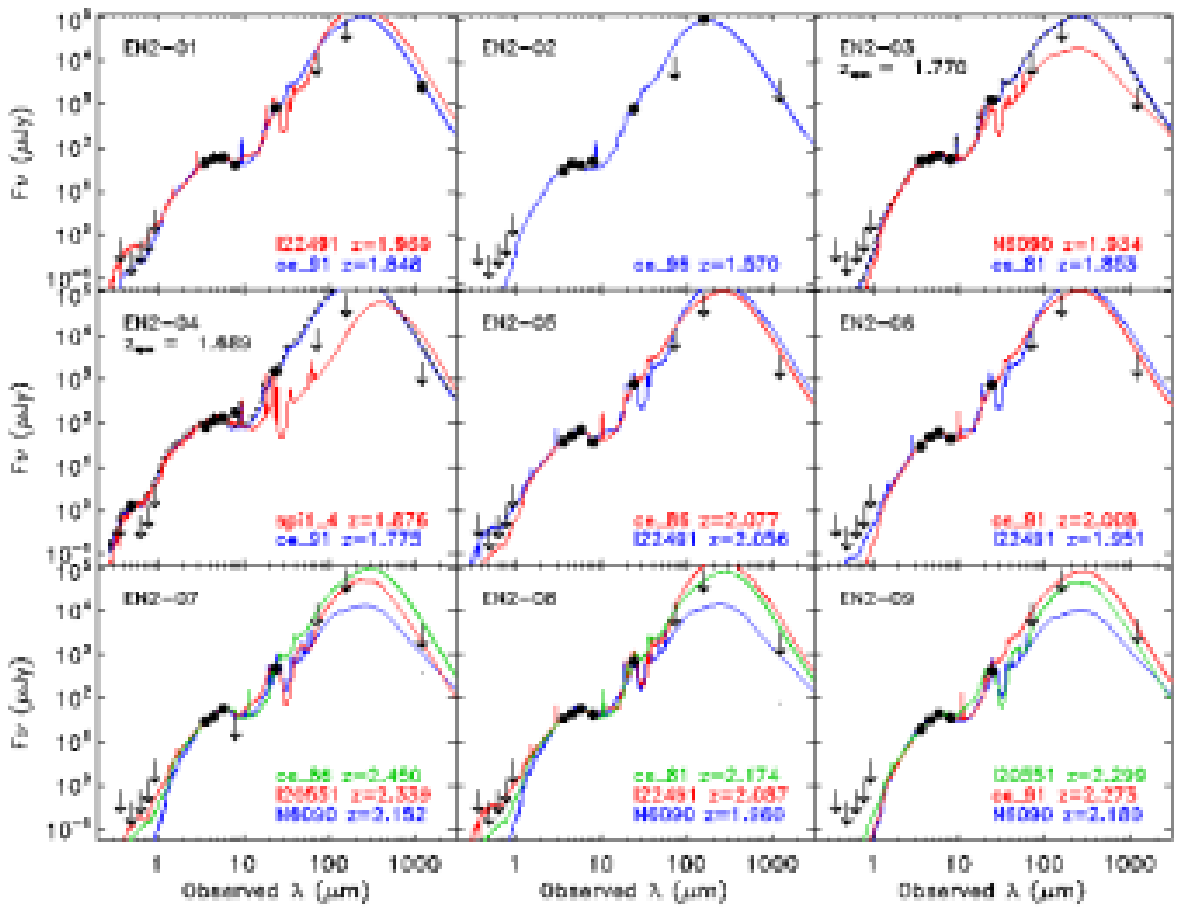}
 \caption{Optical-IR SEDs of MAMBO sources in the ELAIS-N2 field.
  Symbols as in Figure~\ref{sed_fits_zphot_ol}.}
 \label{sed_fits_zphot_o2}
\end{figure*}

\begin{figure*}
\epsscale{1.0}
 \plotone{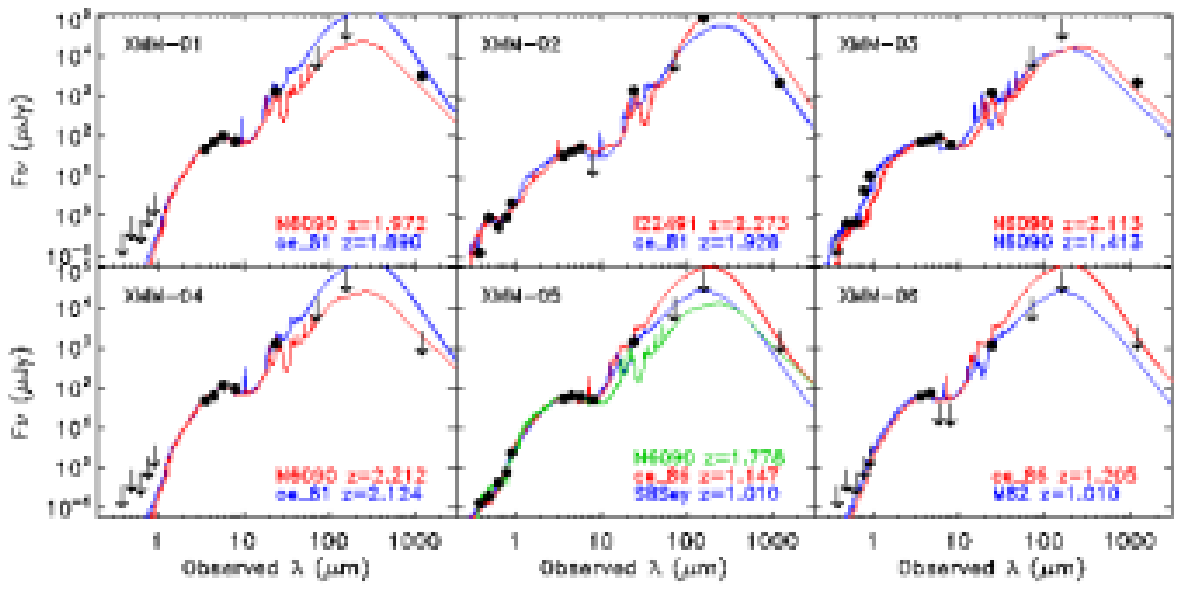}
 \caption{Optical-IR SEDs of MAMBO sources in the XMM-LSS field.
  Symbols as in Figure~\ref{sed_fits_zphot_ol}.}
 \label{sed_fits_zphot_ox}
\end{figure*}



\clearpage
\begin{deluxetable}{lr cccc rrr l}
\tabletypesize{\footnotesize}
\rotate
\tablecaption{Basic properties of the SWIRE-MAMBO sample\label{tab_mm}}
\tablewidth{0pt}
\tablehead{
\colhead{Source ID}&
\colhead{IAU Name\tablenotemark{a}}&
\colhead{$\alpha_{2000}$}&
\colhead{$\delta_{2000}$}&
\colhead{$z$} &
\colhead{$f_{1.2mm}$} &
\colhead{$f_{70}$} &
\colhead{$f_{160}$} &
\colhead{$f_{20cm}$} &
\colhead{Notes\tablenotemark{b}}\\
\colhead{}& 
\colhead{}& 
\multicolumn{2}{c}{(degrees)}& 
\colhead{}&
\colhead{(mJy)}&
\multicolumn{3}{c}{(mJy)} & 
\colhead{}
}
\startdata
\multicolumn{10}{c}{Lockman-Hole}\\                                                                  
\multicolumn{10}{c}{5$\sigma$}\\
   LH-01 & SJ105007.26+571651.0&   162.53026   &    57.280842 &    1.35                    &   5.65$\pm$0.74 &  $<$18 & $<$108 & \nodata &    b3     \\
   LH-02 & SJ103639.57+575346.6&   159.16486   &    57.896271 &    1.93                    &   5.29$\pm$0.85 &  $<$18 & $<$108 & \nodata &    b3     \\
   LH-03 & SJ104313.33+574621.0&   160.80553   &    57.772511 &    2.67                    &   3.79$\pm$0.76 &  $<$18 & $<$108 & \nodata &    b3     \\ 
\multicolumn{10}{c}{4$\sigma$}\\                                                   
   LH-04 & SJ104208.52+582433.3&   160.53551   &    58.409241 &    2.94                    &   2.80$\pm$0.57 &  $<$18 & $<$108 & \nodata &    b3     \\
   LH-05 & SJ103744.45+582950.7&   159.43521   &    58.497410 &    1.88\tablenotemark{c}   &   2.93$\pm$0.61 &  $<$18 & $<$108 & \nodata &    b3/IRS(B5)  \\
   LH-06 & SJ103837.03+582214.7&   159.65428   &    58.370762 &    1.68\tablenotemark{c}   &   3.83$\pm$0.84 &  $<$18 & $<$108 & \nodata &    b2/IRS(B4)  \\
   LH-07 & SJ105405.50+581400.1&   163.52290   &    58.233349 &    1.82\tablenotemark{c}   &   2.76$\pm$0.60 &  $<$18 & $<$108 & \nodata &    b3/IRS(B3)  \\
\multicolumn{10}{c}{3$\sigma$}\\                                                           
   LH-08 & SJ104616.97+580451.9&   161.57072   &    58.081081 &    2.01                    &   2.52$\pm$0.74 &  $<$18 & $<$108 & \nodata &    b3     \\
   LH-09 & SJ104616.96+581659.3&   161.57066   &    58.283150 &    2.28                    &   1.76$\pm$0.56 &  $<$18 & $<$108 & \nodata &    b3     \\
\multicolumn{10}{c}{2$\sigma$}\\                                                   
   LH-10 & SJ103450.43+574116.0&   158.71011   &    57.687771 &    2.23                    &   2.29$\pm$0.77 &  $<$18 & $<$108 & \nodata &    b3/X\tablenotemark{d}     \\
   LH-11 & SJ104601.78+590916.9&   161.50743   &    59.154690 &    1.66                    &   2.06$\pm$0.79 &  $<$18 & $<$108 & 0.082$\pm$0.004 &   b3     \\
   LH-12 & SJ104839.33+592149.0&   162.16386   &    59.363609 &    1.89\tablenotemark{c}   &   2.56$\pm$1.08 &  $<$18 & $<$108 & \nodata &    b4/IRS(B8)  \\ 
\multicolumn{10}{c}{1$\sigma$}\\                                                          
   LH-13 & SJ104656.24+594008.0&   161.73434   &    59.668900 &    1.70                    &   1.92$\pm$0.98 &  $<$18 & $<$108 & \nodata &    b2     \\ 
   LH-14 & SJ104724.25+572150.6&   161.85106   &    57.364052 &    2.03                    &   1.95$\pm$1.03 &  $<$18 & $<$108 & \nodata &    b3     \\
   LH-15 & SJ104822.05+583828.2&   162.09186   &    58.641171 &    2.80                    &   1.13$\pm$0.77 &  $<$18 & $<$108 & \nodata &    b4     \\
   LH-16 & SJ103809.63+591650.1&   159.54013   &    59.280590 &    2.40                    &   1.57$\pm$1.08 &  $<$18 & $<$108 & \nodata &    b4     \\
   LH-17 & SJ105750.75+571652.7&   164.46147   &    57.281300 &    1.85                    &   1.47$\pm$1.10 &  $<$18 & $<$108 & \nodata &    b3     \\
   LH-18 & SJ105943.13+580849.1&   164.92970   &    58.146969 &    2.16                    &   1.46$\pm$1.09 &  $<$18 & $<$108 & \nodata &    b3     \\ 
   LH-19 & SJ104716.02+593201.3&   161.81674   &    59.533691 &    2.11                    &   1.87$\pm$1.58 &  $<$18 & $<$108 & \nodata &    b2/X     \\
   LH-20 & SJ104816.14+592943.3&   162.06725   &    59.495350 &    1.90                    &   1.04$\pm$1.00 &  $<$18 & $<$108 & \nodata &    b2     \\
\multicolumn{10}{c}{$<$1$\sigma$}\\                                                
   LH-21 & SJ103704.32+584755.1&   159.26801   &    58.798641 &    2.63                    &   0.75$\pm$0.90 &  $<$18 & $<$108 & \nodata &    b3     \\
   LH-22 & SJ104646.02+590517.5&   161.69173   &    59.088188 &    1.54                    &   1.40$\pm$1.78 &  $<$18 & $<$108 & 0.051$\pm$0.004 &   b2     \\ 
   LH-23 & SJ103809.17+583226.1&   159.53822   &    58.540581 &    0.98\tablenotemark{c}   &   0.93$\pm$1.17 &  $<$18 & $<$108 & \nodata &    b3/IRS(B6)  \\ 
   LH-24 & SJ103647.44+591819.5&   159.19765   &    59.305408 &    1.25                    &   1.19$\pm$1.57 &  $<$18 & $<$108 & \nodata &    b3     \\
   LH-25 & SJ104059.86+574536.7&   160.24942   &    57.760201 &    2.13                    &   0.74$\pm$1.06 &  $<$18 & $<$108 & \nodata &    b3     \\ 
   LH-26 & SJ103856.94+585244.0&   159.73723   &    58.878891 &    1.88\tablenotemark{c}   &   0.44$\pm$0.93 &  $<$18 & $<$108 & \nodata &    b3/IRS(B7)  \\ 
   LH-27 & SJ104553.76+565930.9&   161.47398   &    56.991909 &    2.01                    &$-$0.10$\pm$1.45 &  $<$18 & $<$108 & \nodata &    b3     \\       
   LH-28 & SJ104942.11+561234.5&   162.42545   &    56.209579 &    2.82                    &$-$0.14$\pm$1.31 &19$\pm$1& $<$108 & 2.73$\pm$0.14 &   b4/X     \\ 
   LH-29 & SJ103500.98+573847.5&   158.75407   &    57.646519 &    2.16                    &$-$1.49$\pm$1.18 &  $<$18 & $<$108 & \nodata &    b3     \\
\multicolumn{10}{c}{ELAIS-N1}\\                                                                  
\multicolumn{10}{c}{3$\sigma$}\\                                                                  
  EN1-01 & SJ160343.08+551735.8&   240.92949   &  55.293282 &    2.37     &   2.81$\pm$0.78 &  $<$17 & $<$104 & \nodata &   b3     \\
  EN1-02 & SJ160440.45+543103.0&   241.16856   &  54.517490 &    1.93     &   2.46$\pm$0.81 &  $<$17 & $<$104 & \nodata &   b3     \\
  EN1-03 & SJ160651.22+545338.0&   241.71341   &  54.893879 &    1.90     &   2.11$\pm$0.69 &  $<$17 & $<$104 & \nodata &   b3     \\
\multicolumn{10}{c}{2$\sigma$}\\                                       
  EN1-04 & SJ161658.14+535319.3&   244.24226   &  53.888691 &    2.00     &   2.82$\pm$0.99 &  $<$17 &81$\pm$4& \nodata &   b3     \\
  EN1-05 & SJ161412.14+541927.2&   243.55058   &  54.324211 &    1.88     &   2.09$\pm$1.03 &  $<$17 & $<$104 & \nodata &   b3     \\
\multicolumn{10}{c}{1$\sigma$}\\                                       
  EN1-06 & SJ160502.00+540639.0&   241.25833   &  54.110840 &    2.37     &   1.15$\pm$0.86 &  $<$17 & $<$104 & \nodata &   b3     \\
  EN1-07 & SJ160409.72+541808.9&   241.04051   &  54.302460 &    1.40     &   1.25$\pm$1.01 &  $<$17 & $<$104 & \nodata &   b3     \\
  EN1-08 & SJ162002.45+542933.8&   245.01019   &  54.492710 &    2.03     &   1.26$\pm$1.02 &  $<$17 & $<$104 & \nodata &   b3     \\
\multicolumn{10}{c}{$<$1$\sigma$}\\                                    
  EN1-09 & SJ161950.53+553740.9&   244.96054   &  55.628040 &    1.69     &   1.07$\pm$1.73 &  $<$17 & $<$104 & \nodata &   b2     \\
  EN1-10 & SJ160945.58+534757.9&   242.43991   &  53.799412 &    2.38     &   0.74$\pm$1.23 &  $<$17 & $<$104 & \nodata &   b4     \\
  EN1-11 & SJ161632.58+540736.9&   244.13573   &  54.126919 &    1.55     &   0.56$\pm$1.04 &  $<$17 & $<$104 & \nodata &   b3     \\
  EN1-12 & SJ161358.92+542812.2&   243.49551   &  54.470051 &    2.11     &   0.04$\pm$1.81 &  $<$17 &55$\pm$4& \nodata &   b3     \\
  EN1-13 & SJ155904.74+551137.2&   239.76977   &  55.193661 &    1.95     &$-$0.11$\pm$1.37 &  $<$17 & $<$104 & \nodata &   b3     \\
  EN1-14 & SJ161753.53+541840.9&   244.47304   &  54.311352 &    2.39     &$-$0.27$\pm$2.00 &  $<$17 & $<$104 & \nodata &   b3     \\
  EN1-15 & SJ161642.78+542355.6&   244.17825   &  54.398769 &    1.96     &$-$0.46$\pm$1.99 &  $<$17 & $<$104 & \nodata &   b3     \\
  EN1-16 & SJ161414.48+554255.4&   243.56035   &  55.715401 &    2.22     &$-$0.99$\pm$2.26 &  $<$17 & $<$104 & \nodata &   b3     \\
  EN1-17 & SJ160611.83+551832.5&   241.54930   &  55.309040 &    2.15     &$-$0.61$\pm$1.38 &  $<$17 & $<$104 & \nodata &   b3     \\
\multicolumn{10}{c}{ELAIS-N2}\\                                                                  
\multicolumn{10}{c}{3$\sigma$}\\                                                                  
  EN2-01 & SJ163734.44+415151.4&   249.393494  &   41.864269  &    1.85                    &   2.52$\pm$0.70 &  $<$18 & $<$124 & \nodata &    b3     \\
\multicolumn{10}{c}{1$\sigma$}\\                                                   
  EN2-02 & SJ163528.58+403843.9&   248.869080  &   40.645519  &    1.57                    &   2.60$\pm$1.47 &11$\pm$1&110$\pm$4& \nodata &    b2     \\ 
  EN2-03 & SJ164058.63+415432.2&   250.244308  &   41.908958  &    1.77\tablenotemark{c}   &   1.15$\pm$0.80 &  $<$18 & $<$124 & \nodata &    fl/IRS     \\ 
  EN2-04 & SJ163022.17+404957.5&   247.592380  &   40.832649  &    1.69\tablenotemark{c}   &   1.17$\pm$0.83 &  $<$18 & $<$124 & \nodata &    b4/IRS    \\
\multicolumn{10}{c}{$<$1$\sigma$}\\                                                
  EN2-05 & SJ163506.01+411735.1&   248.775055  &   41.293072  &    2.06                    &   1.10$\pm$1.16 &  $<$18 & $<$124 & \nodata &    b3     \\
  EN2-06 & SJ163748.16+404922.1&   249.450653  &   40.822811  &    1.95                    &   0.90$\pm$1.21 &  $<$18 & $<$124 & \nodata &    b3     \\
  EN2-07 & SJ163729.80+415722.1&   249.374176  &   41.956150  &    2.15                    &   0.35$\pm$1.55 &  $<$18 & $<$124 & \nodata &           \\
  EN2-08 & SJ163424.64+410954.5&   248.602676  &   41.165131  &    1.96                    &   0.08$\pm$1.07 &  $<$18 & $<$124 & \nodata &    b3     \\
  EN2-09 & SJ163755.89+413416.8&   249.482895  &   41.571320  &    2.18                    &$-$0.24$\pm$2.02 &  $<$18 & $<$124 & \nodata &    b3     \\
\multicolumn{10}{c}{XMM-LSS}\\                                                                  
\multicolumn{10}{c}{4$\sigma$}\\                                                                  
  XMM-01 & SJ021926.24-045212.9&    34.85934  &     -4.870250 &   1.89    &   3.20$\pm$0.71 &  $<$24 &  $<$126 & \nodata &   b3     \\
\multicolumn{10}{c}{3$\sigma$}\\    
  XMM-02 & SJ022521.66-053545.7&    36.34024  &     -5.596040 &   1.93    &   2.45$\pm$0.61 &  $<$24 &103$\pm$4& \nodata &   b3     \\
  XMM-03 & SJ021933.90-043335.5&    34.89124  &     -4.559850 &   2.11    &   2.17$\pm$0.71 &  $<$24 &  $<$126 & \nodata &   b3/X     \\
\multicolumn{10}{c}{1$\sigma$}\\    
  XMM-04 & SJ021925.60-032317.4&    34.85665  &     -3.388180 &   2.12    &   1.46$\pm$0.80 &  $<$24 &  $<$126 & \nodata &   b3     \\  
\multicolumn{10}{c}{$<$1$\sigma$}\\ 
  XMM-05 & SJ021947.11-060017.7&    34.94630  &     -6.004930 &   1.01    &   0.78$\pm$0.90 &  $<$24 &  $<$126 & \nodata &   b3     \\
  XMM-06 & SJ022512.20-053119.6&    36.30082  &     -5.522100 &   1.01    &   0.90$\pm$1.10 &  $<$24 &  $<$126 & \nodata &   b2     \\
\enddata
\tablecomments{The sources are ordered by field and signal-to-noise at
1.2\,mm.}
\tablenotetext{a}{SJ stands for SWIRE\,J. SWIRE\,JHHMMSS.ss+DDMMSS.s is the
official IAU source name for sources discovered in the SWIRE fields.}
\tablenotetext{b}{Shape of the IRAC SED (b2: peak at 4.5\,$\mu$m, b3: peak
at 5.8\,$\mu$m, b4: peak at 8.0\,$\mu$m, fl: flat IRAC SED). Sources with
IRS observations are marked with IRS. B\# refers to the source ID
in~\citet{weedman06a}. Sources with X-ray coverage are marked with X.}
\tablenotetext{c}{Spectroscopic $z$ from IRS spectrum. IRS spectra of LH
sources are published in~\citet{weedman06a}, those of EN2 will be published
in a future publication (Lonsdale et al., in prep.).}
\tablenotetext{d}{X-ray source 425 in CLASX catalog~\citep{yang04}, F(0.4--8 keV) = 2.1$\times$10$^{-15}$ erg cm$^{-2}$ s$^{-1}$ and HR = $-$0.7.}
\end{deluxetable}

\begin{deluxetable}{l rrrrr rrrrr}
\tabletypesize{\footnotesize}
\tablecaption{Optical \& infrared data of the SWIRE-MAMBO sample\label{tab_optir}}
\tablewidth{0pt}
\tablehead{
\colhead{Source ID}&
\colhead{u} &
\colhead{\gp} &
\colhead{\rp} &
\colhead{\ip} &
\colhead{z} &
\colhead{$f_{3.6}$} & 
\colhead{$f_{4.5}$} &
\colhead{$f_{5.8}$} &
\colhead{$f_{8.0}$} &
\colhead{$f_{24}$} \\
\colhead{}& 
\multicolumn{5}{c}{(Vega)} &
\multicolumn{5}{c}{($\mu$Jy)} 
}
\startdata                                                                                                   
 LH-01 &  \nodata &    25.41 &    23.75 &    22.97 &  \nodata &     89 &     94 &    108 &  $<$47 &    755 \\
 LH-02 &  \nodata &  $>$25.2 &  $>$24.4 &  $>$23.5 &  \nodata &     41 &     62 &     72 &  $<$47 &    837 \\
 LH-03 &  \nodata &    24.82 &    24.54 &    23.61 &  \nodata &     39 &     52 &     78 &     64 &    721 \\
 LH-04 &  \nodata &  $>$25.2 &    24.06 &  $>$23.5 &  \nodata &     44 &     63 &    100 &     99 &    707 \\
 LH-05 &  \nodata &    24.40 &    23.45 &    22.73 &  \nodata &     77 &     93 &     98 &     78 &   1502 \\
 LH-06 &  \nodata &    25.50 &  $>$24.4 &    23.61 &  \nodata &     68 &     92 &     97 &     72 &   1072 \\
 LH-07 &  \nodata &    25.23 &    24.68 &    23.98 &  \nodata &     37 &     47 &     62 &     42 &   1203 \\
 LH-08 &  \nodata &  $>$25.2 &  $>$24.4 &  $>$23.5 &  \nodata &     65 &     84 &    116 &     61 &    515 \\
 LH-09 &  \nodata &    24.86 &    24.23 &    23.46 &  \nodata &     60 &     76 &     95 &     85 &    832 \\
 LH-10 &  \nodata &    26.39 &    25.84 &    24.77 &  \nodata &     28 &     45 &     65 &  $<$47 &    484 \\
 LH-11 &  $>$24.3 &    25.05 &    23.70 &    23.17 &    22.36 &     70 &     91 &     95 &     70 &   1034 \\
 LH-12 &  \nodata &  \nodata &    22.62 &  \nodata &  \nodata &     75 &     95 &    120 &    125 &   1419 \\
 LH-13 &    24.20 &    24.83 &  $>$24.4 &  $>$23.5 &  \nodata &     50 &     67 &     57 &  $<$47 &    755 \\
 LH-14 &  \nodata &    25.33 &    24.36 &    23.24 &  \nodata &     42 &     53 &     84 &  $<$47 &    795 \\
 LH-15 &  $>$24.3 &  $>$25.2 &    23.97 &    23.33 &  \nodata &     54 &     75 &    114 &    123 &    638 \\
 LH-16 &  \nodata &  $>$25.2 &  $>$24.4 &  \nodata &  \nodata &     36 &     46 &     62 &     81 &   1318 \\
 LH-17 &  \nodata &  \nodata &    24.88 &    23.81 &  \nodata &     81 &    102 &    104 &     68 &    982 \\
 LH-18 &  \nodata &  $>$25.2 &  $>$24.4 &  $>$23.5 &  \nodata &     20 &     28 &     59 &  $<$47 &    967 \\
 LH-19 &  $>$24.3 &  $>$25.2 &  $>$24.4 &  $>$23.5 &  \nodata &     33 &     40 &     57 &  $<$47 &    488 \\
 LH-20 &  $>$24.3 &  $>$25.2 &  $>$24.4 &  $>$23.5 &  \nodata &     35 &     44 &     48 &  $<$47 &    474 \\
 LH-21 &  \nodata &  $>$25.2 &    23.58 &  $>$23.5 &  \nodata &     54 &     77 &    103 &    104 &    733 \\
 LH-22 &  $>$24.3 &  $>$25.2 &  $>$24.4 &  $>$23.5 &  $>$23.6 &     31 &     40 &     38 &     36 &    446 \\
 LH-23 &  \nodata &  $>$25.2 &  $>$24.4 &  $>$23.5 &  \nodata &     26 &     33 &     40 &  $<$47 &   1111 \\
 LH-24 &  \nodata &    24.77 &    24.07 &    23.00 &  \nodata &     75 &     88 &    103 &     81 &    899 \\
 LH-25 &  \nodata &  $>$25.2 &  $>$24.4 &  $>$23.5 &  \nodata &     31 &     42 &     51 &     46 &    430 \\
 LH-26 &  \nodata &  $>$25.2 &  $>$24.4 &  $>$23.5 &  \nodata &     26 &     45 &     47 &  $<$47 &   1105 \\
 LH-27 &  \nodata &  $>$25.2 &    23.51 &  $>$23.5 &  \nodata &     43 &     54 &     64 &  $<$47 &    673 \\
 LH-28 &  \nodata &  \nodata &    23.65 &  \nodata &  \nodata &     30 &     40 &     46 &     53 &    499 \\
 LH-29 &  \nodata &  $>$25.2 &  $>$24.4 &  $>$23.5 &  \nodata &     31 &     39 &     58 &  $<$47 &    451 \\
\medskip\\
\hline                                                                                                       
\medskip\\
EN1-01 &  $>$23.4 &  $>$24.94 & $>$24.04 &  $>$23.18  &  $>$21.9 &  35 &     48 &     63 &     59 &    698 \\
EN1-02 &  $>$23.4 &  $>$24.94 & $>$24.04 &  $>$23.18  &  $>$21.9 &  37 &     53 &     68 &     44 &    771 \\
EN1-03 &  $>$23.4 &  $>$24.94 & $>$24.04 &  $>$23.18  &  $>$21.9 &  41 &     52 &     57 &     46 &    823 \\
EN1-04 &  $>$23.4 &     24.42 & $>$24.04 &  $>$23.18  &  $>$21.9 &  39 &     49 &     77 &     37 &    657 \\
EN1-05 &  $>$23.4 &  $>$24.94 & $>$24.04 &     23.14  &  $>$21.9 &  30 &     40 &     53 &  $<$47 &    607 \\
EN1-06 &  $>$23.4 &  $>$24.94 & $>$24.04 &  $>$23.18  &  $>$21.9 &  27 &     37 &  $<$40 &     53 &    458 \\
EN1-07 &  $>$23.4 &  $>$24.94 & $>$24.04 &  $>$23.18  &  $>$21.9 &  54 &     67 &     62 &  $<$47 &    560 \\
EN1-08 &  $>$23.4 &  $>$24.94 & $>$24.04 &  $>$23.18  &  $>$21.9 &  19 &     32 &     44 &     34 &    628 \\
EN1-09 &  $>$23.4 &  $>$24.94 & $>$24.04 &  $>$23.18  &  $>$21.9 &  70 &     93 &     88 &     56 &    581 \\
EN1-10 &  $>$23.4 &  $>$24.94 & $>$24.04 &  $>$23.18  &  $>$21.9 &  24 &     37 &     56 &     51 &    446 \\
EN1-11 &  $>$23.4 &  $>$24.94 & $>$24.04 &  $>$23.18  &  $>$21.9 &  37 &   $<$8 &     51 &     67 &    673 \\ 
EN1-12 &  $>$23.4 &  $>$24.94 & $>$24.04 &  $>$23.18  &  $>$21.9 &  44 &     53 &     80 &     63 &    879 \\
EN1-13 &  $>$23.4 &  $>$24.94 & $>$24.04 &  $>$23.18  &  $>$21.9 &  35 &     44 &     51 &     46 &    694 \\
EN1-14 &  $>$23.4 &  $>$24.94 & $>$24.04 &  $>$23.18  &  $>$21.9 &  35 &     46 &     75 &     72 &    550 \\
EN1-15 &  $>$23.4 &  $>$24.94 & $>$24.04 &  $>$23.18  &  $>$21.9 &  33 &     42 &     55 &  $<$47 &    631 \\
EN1-16 &  $>$23.4 &  $>$24.94 & $>$24.04 &  $>$23.18  &  $>$21.9 &  28 &     35 &     48 &  $<$47 &    546 \\
EN1-17 &  $>$23.4 &  $>$24.94 & $>$24.04 &  $>$23.18  &  $>$21.9 &  26 &     41 &     65 &     43 &    599 \\
\medskip\\
\hline
\medskip\\
EN2-01 &  $>$23.4 &  $>$24.94 & $>$24.04 &  $>$23.18  &  $>$21.9 &  45 &     58 &     59 &     40 &    814 \\
EN2-02 &  $>$23.4 &  $>$24.94 & $>$24.04 &  $>$23.18  &  $>$21.9 &  36 &     53 &     48 &     58 &    928 \\
EN2-03 &  $>$23.4 &  $>$24.94 & $>$24.04 &  $>$23.18  &  $>$21.9 &  49 &     54 &     67 &     53 &   1237 \\
EN2-04 &  $>$23.4 &     23.70 & $>$24.04 &  $>$23.18  &  $>$21.9 &  83 &    115 &    129 &    173 &   1521 \\
EN2-05 &  $>$23.4 &  $>$24.94 & $>$24.04 &  $>$23.18  &  $>$21.9 &  37 &     51 &     68 &     37 &    765 \\
EN2-06 &  $>$23.4 &  $>$24.94 & $>$24.04 &  $>$23.18  &  $>$21.9 &  29 &     48 &     60 &     43 &    733 \\
EN2-07 &  $>$23.4 &  $>$24.94 & $>$24.04 &  $>$23.18  &  $>$21.9 &  29 &     38 &     56 &  $<$45 &    472 \\
EN2-08 &  $>$23.4 &  $>$24.94 & $>$24.04 &  $>$23.18  &  $>$21.9 &  34 &     44 &     57 &     42 &    685 \\
EN2-09 &  $>$23.4 &  $>$24.94 & $>$24.04 &  $>$23.18  &  $>$21.9 &  19 &     31 &     44 &     35 &    421 \\
\medskip\\
\hline
\medskip\\
XMM-01 &  $>$25.0 &  $>$24.0 &  $>$24.2 &  $>$24.2 &  $>$23.6 &     47 &     72 &     99 &     74 &   1277 \\
XMM-02 &    26.08 &    23.86 &    24.45 &    23.84 &    22.98 &     36 &     47 &     58 &  $<$66 &   1512 \\
XMM-03 &  $>$25.0 &    24.39 &    24.36 &    22.32 &    21.41 &     70 &     77 &     93 &     58 &   1194 \\
XMM-04 &  \nodata &  \nodata &  $>$24.2 &  \nodata &  \nodata &     47 &     63 &    113 &     95 &   1253 \\
XMM-05 &    26.12 &    25.70 &    24.86 &    24.20 &    22.98 &     54 &     62 &     59 &     50 &   1398 \\
XMM-06 &  $>$25.0 &  $>$24.0 &  $>$24.2 &  $>$24.2 &  $>$23.6 &     62 &     71 &  $<$58 &  $<$66 &   1109 \\
\enddata
\tablecomments{Typical uncertainties are $\sim$5\% for the optical and the
IRAC data, and 10\% for MIPS. Upper limits correspond to 90\% completeness
in the optical and to 5$\sigma$ in the infrared.}
\end{deluxetable}
  
\begin{deluxetable}{lcrrcccc}
\tabletypesize{\footnotesize}
\tablecaption{Statistical information on MAMBO 1.2\,mm results\label{tab_stat}}
\tablewidth{0pt}
\tablehead{
\colhead{Sample}&
\colhead{N}&
\multicolumn{3}{c}{N}&
\colhead{$<f_{1.2mm}(All)>$} &
\colhead{$<f_{1.2mm}(\geq2\sigma)>$\tablenotemark{a}}&
\colhead{$<f_{1.2mm}(<2\sigma)>$\tablenotemark{b}}\\
\colhead{}& 
\colhead{}&
\colhead{$\geq$2$\sigma$}&
\colhead{$\geq$3$\sigma$}&
\colhead{$\gtrsim$4mJy} &
\colhead{(mJy)}&
\colhead{(mJy)}&
\colhead{(mJy)}  
}
\startdata
Total        &   61       & 21 (34\%)            &   16         &     3       &    1.49$\pm$0.18   &   2.90$\pm$0.22      &   0.75$\pm$0.14      \\
LH           &   29       & 12 (41\%)            &    9         &     3       &    1.88$\pm$0.28   &   3.19$\pm$0.36      &   0.95$\pm$0.22      \\
EN1+EN2+XMM  &   32       &  9 (28\%)            &    7         &     0       &    1.14$\pm$0.20   &   2.51$\pm$0.12      &   0.60$\pm$0.17      \\
\enddata
\tablecomments{Uncertainties correspond to the standard deviation of the mean.}
\tablenotetext{a}{Mean 1.2\,mm flux density of sources with $f_{1.2\,mm}>2\sigma$.}
\tablenotetext{b}{Mean 1.2\,mm flux density of sources with $f_{1.2\,mm}<2\sigma$.}
\end{deluxetable}

\begin{deluxetable}{lcc r}
\tabletypesize{\footnotesize}
\tablecaption{Best-fit templates and photometric redshifts\label{tab_fits}}
\tablewidth{0pt}
\tablehead{
\colhead{Source ID}&
\colhead{z}&
\colhead{A$_\mathrm{V}$}&
\colhead{Template} 
}
\startdata
 LH-01 &    1.345 &     1.43 &       SBSey   \\
 LH-02 &    1.925 &     1.69 &      ce\_96   \\
  &    1.934 &     1.95 &       N6090   \\
 LH-03 &    2.674 &     1.04 &      I22491   \\
  &    2.332 &     0.26 &      ce\_81   \\
 LH-04 &    2.943 &     1.04 &      I20551   \\
 LH-05 &    1.583 &     0.39 &      ce\_86   \\

  &    1.734 &     1.04 &     spi1\_4   \\
 LH-06 &    1.791 &     0.78 &      ce\_76   \\
  &    1.803 &     1.56 &      I22491   \\
 LH-07 &    2.225 &     1.17 &      I22491   \\
 LH-08 &    2.014 &     0.78 &      ce\_66   \\
  &    2.035 &     1.04 &      Arp220   \\
 LH-09 &    2.283 &     0.91 &      I20551   \\
 LH-10 &    2.234 &     1.43 &      I20551   \\
 LH-11 &    1.661 &     0.52 &     ce\_105   \\
  &    1.848 &     1.43 &    spi2c\_3   \\
 LH-12 &    1.783 &     0.00 &      ce\_81   \\
  &    2.652 &     0.65 &      I22491   \\
 LH-13 &    1.699 &     0.26 &      ce\_96   \\
 LH-14 &    2.032 &     1.17 &      I22491   \\
  &    2.158 &     0.39 &      ce\_81   \\
 LH-15 &    2.804 &     0.39 &      ce\_81   \\
  &    3.235 &     0.78 &      I20551   \\
 LH-16 &    2.406 &     0.13 &      I19254   \\
  &    2.450 &     1.95 &      I22491   \\
 LH-17 &    1.850 &     0.52 &       N6090   \\
 LH-18 &    2.164 &     1.95 &       N6090   \\
 LH-19 &    2.111 &     1.82 &       N6090   \\
  &    2.359 &     1.04 &      I20551   \\
 LH-19 &    2.362 &     0.65 &      ce\_86   \\
 LH-20 &    1.896 &     1.43 &       N6090   \\
  &    1.990 &     0.52 &      ce\_76   \\
  &    1.945 &     0.52 &      ce\_96   \\
 LH-21 &    2.630 &     0.39 &      ce\_81   \\
  &    2.915 &     0.65 &      I20551   \\
 LH-22 &    1.535 &     1.04 &      ce\_76   \\
  &    1.831 &     0.78 &      I22491   \\
 LH-23 &    1.207 &     1.95 &      ce\_86   \\
 LH-24 &    1.254 &     1.43 &       SBSey   \\
  &    1.756 &     1.17 &      I22491   \\
 LH-25 &    2.133 &     1.43 &      I20551   \\
  &    2.180 &     1.56 &       N6090   \\
 LH-26 &    1.853 &     1.95 &      ce\_81   \\
 LH-27 &    2.014 &     0.52 &      I22491   \\
  &    2.108 &     0.00 &      ce\_81   \\
 LH-28 &    2.815 &     0.65 &      I22491   \\
 LH-29 &    2.164 &     1.69 &       N6090   \\
  &    2.342 &     1.04 &      I20551   \\
\medskip\\
\hline
\medskip\\
 EN1-01 &    2.369 &     1.69 &      I20551  \\
 EN1-02 &    1.934 &     1.95 &       N6090  \\
  &    1.999 &     1.43 &      ce\_91  \\
 EN1-03 &    1.896 &     1.95 &       N6090  \\
  &    2.002 &     0.91 &      ce\_86  \\
 EN1-04 &    2.002 &     0.91 &      I22491  \\
  &    2.136 &     0.13 &      ce\_81  \\
 EN1-05 &    1.879 &     0.26 &      ce\_81  \\
 EN1-06 &    2.366 &     1.95 &       N6090  \\
  &    2.699 &     1.17 &      I20551  \\
 EN1-07 &    1.397 &     1.95 &      ce\_71  \\
  &    1.836 &     0.65 &       N6090  \\
 EN1-08 &    2.026 &     1.95 &       N6090  \\
  &    2.127 &     1.95 &      ce\_81  \\
 EN1-09 &    1.685 &     0.91 &       N6090  \\
  &    1.867 &     0.78 &      ce\_66  \\
 EN1-10 &    2.379 &     1.95 &       N6090  \\
  &    2.744 &     1.30 &      I20551  \\
 EN1-11 &    1.545 &     1.95 &      I20551  \\
  &    3.500 &     0.00 &       N6240  \\
 EN1-12 &    2.111 &     1.95 &       N6090  \\  
  &    2.299 &     1.30 &      ce\_81  \\
 EN1-13 &    1.945 &     1.95 &       N6090  \\
  &    2.164 &     1.04 &      ce\_81  \\
 EN1-14 &    2.392 &     1.95 &       N6090  \\
 EN1-15 &    1.960 &     1.95 &       N6090  \\
  &    2.205 &     0.91 &      ce\_81  \\
 EN1-16 &    2.221 &     1.04 &      ce\_81 \\
  &    2.594 &     1.04 &      I22491 \\
 EN1-17 &    2.152 &     1.95 &       N6090 \\
  &    2.342 &     1.95 &      I20551 \\
\medskip\\
\hline
\medskip\\
 EN2-01 &    1.848 &     0.78 &      ce\_91 \\
  &    1.969 &     1.04 &      I22491 \\
 EN2-02 &    1.570 &     1.95 &      ce\_96 \\
 EN2-03 &    1.853 &     1.69 &      ce\_81 \\
  &    1.934 &     1.95 &       N6090 \\
 EN2-04 &    1.775 &     0.26 &      ce\_91 \\
  &    1.876 &     1.04 &     spi1\_4 \\
 EN2-05 &    2.056 &     1.17 &      I22491 \\
  &    2.077 &     1.04 &      ce\_86 \\
 EN2-06 &    1.951 &     1.95 &      I22491 \\
  &    2.008 &     1.95 &      ce\_91 \\
 EN2-07 &    2.152 &     1.95 &       N6090 \\
  &    2.339 &     1.30 &      I20551 \\
  &    2.450 &     1.04 &      ce\_86 \\
 EN2-08 &    1.960 &     1.95 &       N6090 \\
  &    2.087 &     1.17 &      I22491 \\
  &    2.174 &     1.04 &      ce\_81 \\
 EN2-09 &    2.180 &     1.95 &       N6090 \\
  &    2.273 &     1.95 &      ce\_91 \\
  &    2.299 &     1.95 &      I20551 \\
\medskip\\
\hline
\medskip\\
 XMM-01 &    1.890 &     1.95 &      ce\_81 \\
  &    1.972 &     1.95 &       N6090 \\
 XMM-02 &    1.928 &     0.00 &      ce\_81 \\
  &    2.273 &     0.78 &      I22491 \\
 XMM-03 &    2.113 &     0.00 &       N6090 \\
  &    1.413 &     0.00 &       N6090 \\
 XMM-04 &    2.124 &     1.95 &      ce\_81 \\
  &    2.212 &     1.95 &       N6090 \\ 
 XMM-05 &    1.010 &     1.82 &       SBSey \\
  &    1.147 &     1.17 &      ce\_86 \\
  &    1.778 &     0.13 &       N6090 \\
 XMM-06 &    1.010 &     1.95 &         M82 \\
  &    1.205 &     1.82 &      ce\_86 \\
\hline
\enddata
\tablecomments{The solutions are listed in order of $\chi^2$. All the `ce'
templates are from the library in~\citet{chary01}, the others are from the
library in~\citet{polletta07}. The fits are shown
in Figures~\ref{sed_fits_zphot_ol}, \ref{sed_fits_zphot_o1},
\ref{sed_fits_zphot_o2} and \ref{sed_fits_zphot_ox}.}
\end{deluxetable}

\topmargin=2.cm
\footskip=0in
\begin{deluxetable}{l c c c}
\tabletypesize{\footnotesize}
\tablecaption{FIR and NIR luminosities and dust temperatures for the Lockman Hole sample\label{tab_lum}}
\tablewidth{0pt}
\tablehead{
\colhead{Source}&
\colhead{T$_{dust}$} &
\colhead{$L^{GB}_{FIR}$($z$,$z_{-}$,$z_{+}$)} &
\colhead{$\nu L_{1.6\,\mu m}$} \\
\colhead{ID}&
\colhead{(K)} &
\colhead{(\lsun)} &
\colhead{(\lsun)}
}
\startdata
LH-01 &    26.25  &        12.23 (11.94, 12.46) &   11.93   \\
LH-02 &    31.76  &        12.62 (12.39, 12.79) &   11.69   \\
LH-03 &    38.79  &        12.80 (12.62, 12.93) &   12.28   \\
LH-04 &    41.36  &        12.75 (12.59, 12.87) &   11.91   \\
LH-05 &    31.29  &        12.33 (12.33, 12.33) &   11.74   \\
LH-06 &    29.39  &        12.32 (12.32, 12.32) &   12.07   \\
LH-07 &    30.72  &        12.27 (12.27, 12.27) &   12.09   \\
LH-08 &    32.53  &        12.34 (12.12, 12.51) &   12.57   \\
LH-09 &    35.09  &        12.31 (12.11, 12.47) &   12.15   \\
LH-10 &    34.62  &        12.40 (12.20, 12.56) &   12.25   \\
LH-11 &    29.20  &        12.04 (11.79, 12.23) &   12.03   \\
LH-12 &    31.39  &        12.28 (12.28, 12.28) &   12.04   \\
LH-13 &    29.58  &  $\leq$12.22 (11.97, 12.41) &   12.58   \\
LH-14 &    32.72  &  $\leq$12.44 (12.22, 12.60) &   12.09   \\
LH-15 &    40.03  &  $\leq$12.63 (12.46, 12.75) &   11.96   \\
LH-16 &    36.23  &  $\leq$12.63 (12.44, 12.78) &   12.13   \\
LH-17 &    31.00  &  $\leq$12.36 (12.13, 12.55) &   11.86   \\
LH-18 &    33.95  &  $\leq$12.53 (12.32, 12.69) &   11.79   \\
LH-19 &    33.47  &  $\leq$12.66 (12.45, 12.83) &   12.06   \\
LH-20 &    31.48  &  $\leq$12.35 (12.12, 12.53) &   11.79   \\
LH-21 &    38.42  &  $\leq$12.64 (12.46, 12.77) &   12.21   \\
LH-22 &    28.06  &  $\leq$12.36 (12.10, 12.57) &   11.98   \\
LH-23 &    22.74  &  $\leq$11.63 (11.63, 11.63) &   10.76   \\
LH-24 &    25.31  &  $\leq$12.06 (11.76, 12.30) &   11.54   \\
LH-25 &    33.67  &  $\leq$12.50 (12.29, 12.67) &   12.58   \\
LH-26 &    31.29  &  $\leq$12.31 (12.31, 12.31) &   11.75   \\
LH-27 &    32.53  &  $\leq$12.57 (12.35, 12.75) &   11.92   \\
LH-28 &    40.22  &  $\leq$12.87 (12.70, 12.99) &   11.93   \\
LH-29 &    33.95  &  $\leq$12.56 (12.35, 12.72) &   11.44   \\
\enddata
\tablecomments{$L^{GB}_{FIR}$ is the integrated luminosity between 42.5 and
122.5$\mu$m~\citep{helou88} obtained using a greybody model with
$T_{dust}$=3.9+9.5$\times$(1+$z$), and $\beta$=1.5 (see text). The
luminosities have been obtained at the redshift $z$ of each source. In
brackets are the luminosities derived at $z_{+}=z+0.12\times(1+z)$ and
$z_{-}=z-0.12\times(1+z)$.}
\end{deluxetable}

\begin{deluxetable}{l c cccc}
\tabletypesize{\footnotesize}
\tablecaption{Results from the co-added MIPS images for the Lockman Hole sample\label{tab_coadd}}
\tablewidth{0pt}
\tablehead{
\colhead{Sample}&
\colhead{N}&
\colhead{$<f_{24}>$} &
\colhead{$f_{24}$} &
\colhead{$f_{70}$} &
\colhead{$f_{160}$} \\
\colhead{}&
\colhead{}&
\colhead{detections mean}&
\multicolumn{3}{c}{stack mean}\\
\colhead{}& 
\colhead{}& 
\multicolumn{4}{c}{(mJy)}}
\startdata
All                          & 27 &  0.83$\pm$0.31 & 0.93$\pm$0.08 &    5.7$\pm$1.5  &   39.0$\pm$10.2  \\
$f_{1.2mm}>2\sigma$          & 12 &  0.92$\pm$0.33 & 0.99$\pm$0.11 &    5.7$\pm$1.3  &   57.8$\pm$16.3 \\
$f_{1.2mm}<2\sigma$          & 15 &  0.75$\pm$0.28 & 0.90$\pm$0.11 &    7.3$\pm$2.8  &   16.9$\pm$8.2  \\
\enddata
\end{deluxetable}

\begin{deluxetable}{l cc}
\tabletypesize{\footnotesize}
\tablecaption{Stellar masses for the Lockman Hole sample\label{tab_masses}}
\tablewidth{0pt}
\tablehead{
\colhead{Source ID}&
\colhead{$M_{\star}$} &
\colhead{\av} \\
\colhead{}&
\colhead{(10$^{11}$\,\msun)}&
\colhead{}
}
\startdata
 LH-01 &  1.21$^{+0.07}_{-0.03}$  &  2.32$^{+0.02}_{-0.03}$  \\
 LH-02 &  1.51$^{+0.14}_{-0.06}$  &  2.27$^{+0.05}_{-0.18}$  \\
 LH-03 &  2.53$^{+0.36}_{-0.18}$  &  1.70$^{+0.06}_{-0.10}$  \\
 LH-04 &  4.66$^{+0.19}_{-0.15}$  &  2.51$^{+0.03}_{-0.09}$  \\
 LH-05 &  2.64$^{+0.39}_{-0.03}$  &  1.85$^{+0.02}_{-0.15}$  \\
 LH-06 &  1.71$^{+0.08}_{-0.05}$  &  2.37$^{+0.04}_{-0.06}$  \\
 LH-07 &  1.06$^{+0.05}_{-0.03}$  &  2.30$^{+0.03}_{-0.05}$  \\
 LH-08 &  4.94$^{+0.10}_{-0.56}$  &  1.67$^{+0.02}_{-0.16}$  \\
 LH-09 &  3.25$^{+0.25}_{-0.14}$  &  1.80$^{+0.04}_{-0.09}$  \\
 LH-10 &  1.61$^{+0.10}_{-0.05}$  &  2.13$^{+0.05}_{-0.08}$  \\
 LH-11 &  2.34$^{+0.15}_{-0.04}$  &  1.93$^{+0.03}_{-0.06}$  \\
 LH-12 &  3.14$^{+0.31}_{-0.20}$  &  1.25$^{+0.03}_{-0.04}$  \\
 LH-13 &  1.43$^{+0.29}_{-0.12}$  &  1.76$^{+0.02}_{-0.17}$  \\
 LH-14 &  1.82$^{+0.15}_{-0.08}$  &  1.85$^{+0.04}_{-0.16}$  \\
 LH-15 &  5.60$^{+0.15}_{-0.14}$  &  1.86$^{+0.05}_{-0.05}$  \\
 LH-16 &  1.93$^{+0.17}_{-0.09}$  &  2.73$^{+0.03}_{-0.13}$  \\
 LH-17 &  2.70$^{+0.31}_{-0.15}$  &  2.05$^{+0.04}_{-0.10}$  \\
 LH-18 &  0.85$^{+0.12}_{-0.03}$  &  2.85$^{+0.02}_{-0.20}$  \\
 LH-19 &  2.31$^{+0.23}_{-0.13}$  &  1.74$^{+0.20}_{-0.22}$  \\
 LH-20 &  1.43$^{+0.33}_{-0.09}$  &  1.79$^{+0.16}_{-0.36}$  \\
 LH-21 &  4.37$^{+0.27}_{-0.12}$  &  1.88$^{+0.05}_{-0.06}$  \\
 LH-22 &  0.60$^{+0.04}_{-0.02}$  &  2.49$^{+0.08}_{-0.15}$  \\
 LH-23 &  0.23$^{+0.06}_{-0.00}$  &  2.82$^{+0.02}_{-0.29}$  \\
 LH-24 &  0.90$^{+0.07}_{-0.03}$  &  2.43$^{+0.03}_{-0.04}$  \\
 LH-25 &  1.53$^{+0.25}_{-0.04}$  &  2.17$^{+0.02}_{-0.19}$  \\
 LH-26 &  0.81$^{+0.08}_{-0.02}$  &  2.76$^{+0.03}_{-0.24}$  \\
 LH-27 &  2.22$^{+0.51}_{-0.03}$  &  1.42$^{+0.02}_{-0.21}$  \\
 LH-28 &  1.86$^{+0.27}_{-0.10}$  &  1.76$^{+0.05}_{-0.09}$  \\
 LH-29 &  1.49$^{+0.22}_{-0.06}$  &  1.98$^{+0.16}_{-0.26}$  \\
\enddata
\tablecomments{Uncertainties correspond to 1$\sigma$.}
\end{deluxetable}

\end{document}